\documentclass[11pt,onecolumn,twoside,notitlepage,fleqn,letterpaper,final]
	      {article}[]

\usepackage{amsmath}
\usepackage{amssymb}
\usepackage{latexsym}
\usepackage{amsfonts}
\usepackage[mathscr]{eucal}
\usepackage{array}
\usepackage{calc}
\usepackage{longtable}
\usepackage{multirow}
\usepackage{bm}
\usepackage{slashed}
\usepackage[dvips]{epsfig}
\usepackage{axodraw}
\usepackage{pstricks}
\usepackage{graphicx}
\usepackage{xspace}
\usepackage{cite}
\usepackage{setspace}
\usepackage[T1]{fontenc}
\usepackage{textcomp}
\usepackage[font=small,labelfont=bf]{caption}
\usepackage{sectsty}
\allsectionsfont{\sffamily}
\subsubsectionfont{\bfseries\slshape\large}
\DeclareMathAlphabet{\mathitbf}{OML}{cmm}{b}{it}
\usepackage{fancyhdr}
\let\spreprint\empty
\newcommand{\preprint}[1]{\def\spreprint{\protect#1}}
\let\sinstitute\empty
\newcommand{\institute}[1]{\def\sinstitute{\protect#1}}
\makeatletter
\newcommand{\mymaketitle}{\begingroup
  \null\thispagestyle{empty}%
    \ifx\spreprint\empty
      \vskip5ex
    \else
      \vspace*{-10ex}\flushright\small\spreprint\vskip2ex
    \fi
    \vskip7ex
    \begin{center}
      \doublespacing{\sffamily\bfseries\LARGE\@title}\vskip5ex
      {\bf\large\@author\vskip 2ex}
      \ifx\sinstitute\empty
      \else
        {\sl\small\sinstitute}
      \fi
    \end{center}
    \vskip7ex
  \endgroup
}
\makeatother
\renewenvironment{abstract}{\begin{center}
  {\large\sffamily\bfseries Abstract}\\[4mm]
  \begin{minipage}[t]{0.87\textwidth}\small
}{\end{minipage}\end{center}\vskip10ex}

\newcommand{\bit}{\begin{itemize}}
\newcommand{\eit}{\end{itemize}}
\newcommand{\bc}{\begin{center}}
\newcommand{\ec}{\end{center}}
\newcommand{\bq}{\begin{quote}}
\newcommand{\eq}{\end{quote}}
\newcommand{\bi}{\begin{itemize}}
\newcommand{\ei}{\end{itemize}}
\newcommand{\be}{\begin{equation}}
\newcommand{\ee}{\end{equation}}
\newcommand{\bea}{\begin{eqnarray}}
\newcommand{\eea}{\end{eqnarray}}
\newcommand{\bt}{\begin{tabular}}
\newcommand{\et}{\end{tabular}}
\newcommand{\bmp}{\begin{minipage}}
\newcommand{\emp}{\end{minipage}}
\newcommand{\btab}{\begin{tabbing}}
\newcommand{\etab}{\end{tabbing}}

\newcommand{\beq}{\begin{equation}}
\newcommand{\eeq}{\end{equation}}
\newcommand{\beqa}{\begin{eqnarray}}
\newcommand{\eeqa}{\end{eqnarray}}
\def\trm{\textrm}
\def\nn{\\ \nonumber}
\def\nnb{\nonumber}

\def\mrm{\mathrm}
\def\cal{\mathcal}

\def\eg{\trm{e.g.\ }}
\def\ie{\trm{i.e.\ }}

\def\eq#1{eq.~(\ref{#1})}
\def\Eq#1{Eq.~(\ref{#1})}

\def\Eqs#1{Eqs.~(\ref{#1})}

\def\Figs#1{Figures~\ref{#1}}

\def\Tab#1{Table~\ref{#1}}

\def\Sec#1{Section~\ref{#1}}

\def\App#1{Appendix~\ref{#1}}
\preprint{FERMILAB-PUB-11-200-T}
\author{\hphantom{MisterX}\\[0mm]\sffamily\bfseries\normalsize%
  Walter~T.~Giele\thanks{giele@fnal.gov}
  \,\!,\ \
  Gerben~C.~Stavenga\thanks{stavenga@fnal.gov}
  \\[4mm]
  {\small\sl Theoretical Physics Department}\\[-2mm]
  {\small\sl Fermi National Accelerator Laboratory, Batavia, IL 60510, USA}
  \\[10mm]\sffamily\bfseries\normalsize%
  Jan Winter\thanks{jwinter@cern.ch}
  \\[4mm]
  {\small\sl PH-TH Department, CERN, CH-1211 Geneva 23, Switzerland}
  \\[9mm]
}
\title{\sffamily\bfseries\Large%
  A Forward Branching Phase-Space Generator}
\institute{Theoretical Physics Department, Fermi National Accelerator
  Laboratory, Batavia, IL 60510, USA}

\newcommand{\order}[1]{$\mathcal O(#1)$}

\begin{document}
\renewcommand{\baselinestretch}{1.4}
\maketitle
\thispagestyle{empty}
\begin{flushright}
  \vspace*{-128mm}
  {\small FERMILAB-PUB-11-200-T\\CERN-PH-TH/2011-109}\\[131mm]
\end{flushright}

\renewcommand{\baselinestretch}{1.1}
\begin{abstract}
\noindent
We develop a forward branching phase-space generator for use in
next-to-leading order parton level event generators.
By performing $2\to3$ branchings from a fixed jet phase-space
point, all bremsstrahlung events contributing to the
given jet configuration are generated.
The resulting phase-space integration is three-dimensional
irrespective of the considered jet multiplicity.
In this first study, we use the forward branching phase-space
generator to calculate in the leading-color approximation
next-to-leading order corrections to fully differential gluonic jet
configurations.
\end{abstract}

\clearpage
\renewcommand{\baselinestretch}{1.2}
\tableofcontents
\thispagestyle{empty}
\renewcommand{\baselinestretch}{1.04}
\vspace*{4mm}
\noindent\hrulefill
\vspace*{0mm}


\section{Introduction}\label{sec:Intro}

The evaluation of the virtual corrections to processes including a large number
of jets has become straightforward~\cite{Giele:2008bc,Ellis:2008qc,
Lazopoulos:2008ex,Ellis:2009zw,Winter:2009kd,Berger:2009zg,
Melnikov:2009dn,Bevilacqua:2009zn,Giele:2009ui,Bevilacqua:2010ve,Berger:2010vm,
Melnikov:2010iu,Melia:2010bm,Frederix:2010ne,Berger:2010zx,Badger:2010nx,
Bern:2011ie,Melia:2011dw,Bern:2011pa}
due to the advent of algorithmic implementations~\cite{Ellis:2007br,
Giele:2008ve,Berger:2008sj,Ellis:2008ir,vanHameren:2009dr,
Mastrolia:2010nb,Hirschi:2011pa,Ellis:2011cr}
of generalized unitarity~\cite{Bern:1994cg,Britto:2005ha} using
parametric integration techniques~\cite{Ossola:2006us}.
The only required diagrammatic evaluations are partonic tree-level
amplitudes. These diagrams can be efficiently evaluated through
algorithmic recursion relations~\cite{Berends:1987me,Kosower:1989xy,
Caravaglios:1995cd,Kanaki:2000ey,Moretti:2001zz,Cachazo:2004kj,
Risager:2005vk,Britto:2004ap,Britto:2005fq,Draggiotis:2005wq,
Vaman:2005dt,Schwinn:2005pi}.
As a consequence, all building blocks exist to construct a
Next-to-Leading Order (NLO) 
event generator for the evaluation of high multiplicity jet observables.
Such a generator would expand the possible phenomenological studies
at, for example, the Large Hadron Collider (LHC) significantly.

However, the integration of the bremsstrahlung events
over phase space hinders further development of such
generators. Current methods for the automated numerical 
integration~\cite{Frixione:1995ms,Catani:1996vz,Gleisberg:2007md,
Frederix:2008hu,Czakon:2009ss,Frederix:2009yq,Hasegawa:2009tx,Frederix:2010cj}
of the real matrix elements over the high dimensional phase space
require large computer resources. Already for $2\to4$ processes, the
obtainable statistical accuracy is a limiting factor for realistic
phenomenology.

On the other hand, shower Monte Carlo
programs~\cite{Corcella:2000bw,Bahr:2008pv,Sjostrand:2006za,Sjostrand:2007gs,
Gleisberg:2003xi,Gleisberg:2008ta,Buckley:2011ms}
tell us efficient algorithms exist to generate events closely
following the physics \cite{Bauer:2008qj}. It is therefore of interest
to investigate phase-space generators based on especially dipole
showers~\cite{Lonnblad:1992tz,Sjostrand:2004ef,Giele:2007di,Dinsdale:2007mf,
Schumann:2007mg,Winter:2007ye,Larkoski:2009ah,Platzer:2009jq}
as they employ exact phase-space factorization. A shower based NLO
phase-space generator uses as a starting point an $n$-parton final
state, which can have additional multiple non-partonic
particles.\footnote{At this early stage of developing our method, we do
  not consider the cases where partonic decays of color-neutral bosons
  may contribute to the $n$-parton final state.}
The origin of this final state does not affect the NLO generator. It
can be a previously generated Leading Order (LO) event (weighted or
unweighted), an event provided by a phase-space generator such as
\textsc{Rambo}~\cite{Kleiss:1985gy}, \textsc{Sarge}~\cite{Draggiotis:2000gm},
\textsc{Haag}~\cite{vanHameren:2002tc}, \textsc{Kaleu}~\cite{vanHameren:2010gg}
or any other source. The Forward Branching Phase-Space (FBPS)
generator will interpret each of the $n$\/ parton momenta as a jet
momentum. Next, following the dipole shower formalism, it performs
forward $2\to3$ branchings, thereby generating the $(n+1)$-parton phase
space. The brancher is constructed in a way that a 
jet algorithm based on a $3\to2$ clustering algorithm
exactly inverts the branching. As a consequence the
observed final state is unaltered during the phase-space integration.

As a result, any observable constructed from the {\em fixed}\/ number of
jet momenta remains {\em unaltered}\/ by the FBPS generator and hence
does not constrain the real-emission phase-space integration, which
can thus be viewed as integrating out the partonic degrees of freedom
inside the jets. In this sense the jet has become opaque and no
information can be extracted about the internal jet structure, which
is the domain of resummed calculations, in particular parton showers.
We note that if a dipole shower would use the same brancher, the
matching of this shower to the NLO calculation becomes a triviality as
the shower, just as the observables, factorizes from the
bremsstrahlung integration. In other words, the shower does not alter
the NLO event weight nor does it change the partonic jet observable.

The FBPS generator is an one-particle phase-space integrator, independent from 
the number of jets. Therefore, for a given $n$-jet configuration, the 
numerical accuracy is only
affected by the number of possible $2\to3$ branchings, \ie the number of jet pairs. 
This means that in order to maintain constant statistical accuracy, as 
the number of jets increase,
the Monte Carlo program needs \order{N_\mrm{evts}\times n^2}
generated bremsstrahlung events where $N_\mrm{evts}$ is a number of
events independent of the jet multiplicity $n$.
An additional virtue is that all generated bremsstrahlung events 
are added to the same virtual event, making
the infra-red/collinear cancellations efficient and easy to optimize 
in the three-dimensional phase space.
This allows us to use a simple
slicing method to facilitate the cancellation of the 
infra-red/collinear singularities. Note that
subtraction methods can be trivially implemented as any 
jet observable does not depend on the generated
bremsstrahlung event.

In the first implementation of this method, we use a special $3\rightarrow 2$
clustering jet algorithm, which is an augmented $2\rightarrow 1$ jet algorithm. 
This augmentation adds a recoil parton to the $2\rightarrow 1$ clustering.
As a result, the NLO jet phase space becomes identical 
to the LO jet phase space. Specifically,
jets resulting from the clustering remain 
massless and the jet algorithm preserves
momentum conservation, \ie particles ``clustered'' with the 
beam are not discarded. From these
observations it is clear, the jet augmentations will only modify the
last step in the jet algorithm: the clustering prescription. The
clustering step is added rather ad-hoc to current jet algorithms,
which makes it easier to implement the necessary modifications as
described later on in this paper.

The layout of the paper is as follows. In \Sec{sec:Jets} we discuss
issues regarding the infra-red safety of jet observables. We will
propose augmentations of the jet algorithms to make all possible jet
observables infra-red safe. Because of their enhanced infra-red safe
behavior, these augmented jet algorithms have improved theoretical
properties.
We will use these properties to construct the FBPS generator in
\Sec{sec:FBPS}. With this phase-space generator at hand, we build --
as a proof-of-principle -- a leading-color NLO parton level event
generator for $PP\to n$\/ gluonic jets. This is detailed in
\Sec{sec:POP}, before we give our conclusions in \Sec{sec:Conclusions}. 
Finally, two appendices are added. The first appendix details the
number and type of branchings occurring in the FBPS generator.
The second appendix lists the explicit jet
configurations utilized in \Sec{sec:POP} to perform the numerical studies.


\section{The Fully Differential Jet Cross Section}
\label{sec:Jets}

At hadron colliders, jets form the basis of defining the event
topology and thereby characterize the underlying hard scattering
event. It is therefore imperative to understand jets in both experiment
and theory. An essential requirement of the jet defining algorithm is
infra-red finiteness, which expresses the fact that the addition of
arbitrary soft/collinear particles does not alter the final-state
jet multiplicity. The
infra-red finiteness requirement in jet algorithms is by now well understood
(see \eg~\cite{Cacciari:2008gp}).

Before the advent of the numerical parton level NLO
generators~\cite{Giele:1991vf,Giele:1993dj}, 
semi-numerical programs calculated corrections to differential
jet observables. For example, in Ref.~\cite{Ellis:1994dg} the NLO
corrections to the semi-exclusive dijet cross section are calculated
for explicitly given values of the dijet mass and the rapidities of
the two leading jets. This gives a necessarily infra-red safe (finite)
correction for each point in the dijet phase space. Note that the jet
algorithm inevitably forms an integral part of this calculation.

Current parton level NLO multi-jet generators perform a Monte Carlo
integration over the brems\-strah\-lung phase space independently of
any actual jet algorithm. This has the apparent advantage that any
experimental jet algorithm can be numerically accommodated in the
Monte Carlo programs. However, owing to the infra-red properties of the jet
algorithms, the generators only produce infra-red safe results for more
inclusive jet observables. The resulting predictions from the Monte
Carlo generators are not infra-red safe for each point in the multi-jet phase
space as the LO and NLO jet phase spaces only coincide on the boundary
defined by soft/collinear emissions. A sufficient amount of jet
phase-space averaging is required to obtain finite results. For
example, one cannot use arbitrarily small bin sizes in representing
the results of the Monte Carlo integration -- sufficiently wide bins
are needed.

As mentioned above while current jet algorithms are infra-red finite,
the observables constructed from these jets are not necessarily
infra-red safe. This is a direct
consequence of the clustering procedure constituting the last step of
the jet algorithms. For example, the dijet azimuthal angle
de-correlation is a typical non infra-red safe jet
observable~\cite{Abazov:2004hm,Khachatryan:2011zj}.
At LO ($n=2$), the two jets in the event are exactly back-to-back in
the azimuthal plane. At NLO, the generation of the jet mass will cause
the bremsstrahlung events to deviate from the back-to-back
configuration leaving uncanceled logarithmic divergences. On the
contrary, with an infra-red safe jet algorithm, the NLO jet phase space is
identical to the LO jet phase space and the two jets remain balanced.
All that is calculated at NLO is the $K$-factor; only a third jet will induce
a de-correlation.

It is important to note that 
the Kinoshita--Lee--Nauenberg (KLN) theorem is not a jet
phase-space averaged property. By integrating over all partonic
contributions to a {\em fixed}\/ jet configuration, the KLN theorem
should already hold. In other words, an infra-red safe jet algorithm
must provide a proper cancellation between virtual and bremsstrahlung
events for {\em each}\/ jet phase-space point. As a consequence any jet
observable constructed through an infra-red safe jet algorithm is
finite.

\subsection{Jet algorithms and infra-red safe observables}
\label{subsec:algodef}

To explore the issues with jet algorithms further, we will
first look at final-state jets at a lepton collider:
$\ell^+\ell^-\to J_1\cdots J_n$. Here, a sequential jet algorithm is
readily constructed by defining, as a function of the cluster momenta $\{c_i\}$,
an event resolution measure $R_\mrm{evt}(c_1,\ldots,c_m)$.
The event resolution separates the soft/collinear region from the hard region 
of phase space. The clusters can be individual hadrons,
calorimeter cells, tracks and combinations thereof.
If $R_\mrm{evt}$ is smaller than the requested jet resolution,
$R_\mrm{jet}$, the number of clusters is recombined using the 
cluster procedure defined by the specific jet algorithm:
$\{c_1,\ldots,c_m\}\mapsto\{\hat{c}_1,\ldots,\hat{c}_{m-1}\}$.
This is repeated until $R_\mrm{evt}\geq R_\mrm{jet}$, at which point
the remaining clusters are identified as the jets.

Jet algorithms currently used by experiments (\cite{Catani:1993hr,
  Ellis:1993tq,Blazey:2000qt,Salam:2007xv,Ellis:2007ib,Cacciari:2008gp})
define the event resolution function in terms of resolution functions
of pairs of clusters: $R_\mrm{evt}=\min_{ij} R_{ij} : R_{ij}=R(c_i,c_j)$.
The minimization procedure identifies the least resolved pair of clusters and
recombines these two clusters to one new cluster, thereby decreasing
the total number of clusters in the event by one. In here lies a fundamental
issue: either the newly formed cluster has a four-momentum, which is
massive due to adding the momenta of the two clusters $\hat{c}_{ij}=c_i+c_j$,
or overall momentum conservation is violated.
From a theoretical point of view, the $2\to1$ clustering causes the
NLO jet phase space to separate from the LO jet phase space. The LO
jets are massless, while the NLO jets now necessarily are
massive. They only match in the exact soft/collinear limit, in which
case the new cluster is massless: $\hat{c}_{ij}=0$.
As a result we have infra-red finiteness, but no infra-red safety. The
necessary care has to be taken when defining jet observables. For a
given value of a jet observable ${\cal O}_\mrm{obs}$, the virtual
correction contributes at a single point $\delta({\cal O}-{\cal O}_\mrm{obs})$
while, because of the jet mass, the bremsstrahlung is distributed as
$\log({\cal O}-{\cal O}_\mrm{obs})$. This behavior is ``cured'' by
allowing, for example, sufficient smearing in histogram bins.

To maintain infra-red safety, we need to both keep massless
clusters and maintain overall momentum conservation when combining
clusters. The minimal procedure to do this is by defining the
resolution function in terms of triplets of clusters~\cite{Lonnblad:1992qd}:
$R_\mrm{evt}=\min_{ijk} R(c_i,c_j,c_k)$. The triplets of clusters
$\{c_i,c_j,c_k\}$ can be recombined to pairs of clusters
$\{\hat{c}_i,\hat{c}_j\}$, while maintaining both momentum
conservation, $\hat{c}_i+\hat{c}_j=c_i+c_j+c_k$, and
keeping the newly formed clusters massless,
$\hat{c}_i^2=\hat{c}_j^2=0$.\footnote{If flavor-tagged clusters are
  involved, $\hat{c}_i^2\neq0$, the possible quark mass has to be
  taken into consideration.}
With this type of jet algorithm one can define infra-red safe jet
cross sections. As a consequence the fully differential cross section 
$d^{(n)}\sigma/d\,J_1\cdots d\,J_n$, and all possible distributions
of jet observables derived from it, are infra-red safe. The
reason this can be done is that the LO and NLO jet phase spaces exactly match.
We can therefore construct a phase-space brancher similar to the ones used in 
dipole showers~\cite{Giele:2007di,Winter:2007ye}.
By choosing the branching map as the inverse of the $3\to2$ clustering
used in the jet algorithm, all generated bremsstrahlung events are
mapped back to the same jet phase-space point. This results in an
infra-red safe, fully differential jet cross section by virtue of the
KLN theorem. That is, both virtual and bremsstrahlung corrections contribute
to $\delta({\cal O}-{\cal O}_\mrm{obs})$ only and no smearing is required
to obtain a finite result.

As is clear from the above discussions, it is straightforward to
construct a FBPS generator for lepton colliders. It would calculate
the $K$-factor to a fixed jet phase-space point, \ie the fully
differential jet cross section. However, for hadron colliders, the
incoming partons cause additional complications. The current jet
algorithms used in hadron colliders augment the lepton collider jet
algorithms by including a resolution measure of clusters with respect
to the beam. If a cluster is combined with the beam, it is effectively
removed. As a result the remaining clusters violate momentum
conservation as we have un-clustered momenta. To get infra-red safety,
momentum conservation must be preserved during the clustering. There
are two options: either build up a beam jet, or perform final-state
clusterings {\em only}.

The first option is to construct a beam jet: instead of removing the
final-state clusters, when combining with the incoming beam, they are
combined with the respective (separate) beam cluster. Once the event
resolution passes the jet resolution, we are left with two incoming
beam jets and the final-state jets. All jets are massless and
four-momentum is conserved. However, the two beam-jet momenta are not
along the incoming (anti-)proton directions. To map onto the LO jet
phase space, where the two beam-jet momenta {\em are along}\/ the
incoming (anti-)proton directions, we have to define the jet
observables in the frame where the two beam jets are along the
(anti-)proton directions. That is, we have to perform a transverse
momentum boost to this frame. From a theoretical point of view, this
has the desirable feature that the effect of ``initial-state
radiation'' is minimized as this radiation does not affect the
observable due to the boost. Effectively, the initial-state radiation
is integrated out within the jet cone and the KLN theorem guarantees a
properly defined fully differential jet cross section.

The second option is to constrain the initial-state clusters to remain
along the respective beam direction during the clustering phase: the
beam particle momenta are only rescaled. While not immediately
obvious, this can always be accomplished using the $3\to2$ clustering
maps.\footnote{This clustering only works for final states with at
  least one jet.} From an experimental point of view, this is not a
particular desirable option as all radiation is assigned to the
final-state jets.

For a proof-of-principle calculation, the second option is highly
desirable as it minimizes theoretical complications. It will be used
in this paper. As this is a NLO calculation, only one clustering step
is performed. We start with the partonic scattering $p_a\,\!p_b\to
p_1\cdots p_{n+1}$ and reduce this to the jet final state $J_a\,\!J_b\to
J_1\cdots J_n$. Note that $a$\/ and $b$\/ are only used to label the
incoming partons or jets; no flavor information is associated with
these labels. Out of the large class of infra-red safe jet algorithms,
which can be constructed, the explicit jet algorithm used in this
paper is as follows:
\begin{enumerate}
\item Find initial- or final-state parton $i$\/ and final-state parton
  $j$\/ by minimizing the resolution parameter
  \be\label{eq:step1}
  R_{ij}\;=\;|\,\!s_{ij}|\,=\,\left| (\pm p_i+p_j)^2\right|
  \ee
  where ``$+$'' is used for $i$\/ being a final-state particle and
  ``$-$'' for being an initial-state particle.
\item Given partons $i$\/ and $j$\/ of the previous step, find
  final-state parton $k$\/ by minimizing
  \be\label{eq:step2}
  R_{ij;k}\;=\;\min\left(R_{ik},R_{jk}\right)\ .
  \ee
\item If parton $i$\/ is a final-state parton:\quad
  cluster parton $i$ and $j$, $p_{ij}=p_i+p_j$, and
  use parton $k$\/ as the recoil momentum to make the cluster
  massless:
  \beq\label{eq:step3}
  \left\{\begin{array}{l}
  J_i\;=\;p_{ij}+(1-\gamma)\,p_k\ ,\\[2mm]
  J_k\;=\;\gamma\,p_k
  \end{array}\right.
  \eeq
  with $\gamma=1+s_{ij}/(s_{ik}+s_{jk})$. This maps the three
  final-state partons onto two massless jets:
  $\{p_i,p_j,p_k\}\mapsto\{J_i,J_k\}$ while preserving momentum
  conservation: $J_i+J_k=p_i+p_j+p_k$.
\item If parton $i$\/ is an initial-state parton, say $i=a$:\quad
  cluster the two final-state momenta, $p_{jk}=p_j+p_k$, and use the
  initial-state parton $a$\/ as the recoil momentum to make the
  cluster massless:
  \beq\label{eq:step4}
  \left\{\begin{array}{l}
  J_a\;=\;\gamma\,p_a\ ,\\[2mm]
  J_j\;=\;p_{jk}-(1-\gamma)\,p_a
  \end{array}\right.
  \eeq
  with $\gamma=1-s_{jk}/(s_{aj}+s_{ak})$. The two final-state partons
  are now mapped onto one massless jet and a rescaled initial-state
  parton: $\{p_a,p_j,p_k\}\mapsto\{J_a,J_j\}$, while maintaining
  overall momentum conservation $J_a-J_j=p_a-p_j-p_k$.
\end{enumerate}

We will use an {\em inclusive}\/ mode of the algorithm. This means, we
keep clustering until the desired (LO predefined) number of jets is
reached. The alternative is to cluster until the jet resolution
exceeds the preset minimum, after which the event is vetoed, if the
number of jets is not equal to the desired number of jets. This is the
{\em exclusive}\/ mode of the algorithm, which is {\em not}\/ used in
this paper. Note that for reproducing the usual NLO $n$-jet inclusive
observables, we have to perform a two-stage run. First, generate the
NLO $K$-factors for the exclusive $n$-jet events, next, add the
$(n+1)$-jet events at LO. From this event sample the observable can be
determined.

\subsection{Defining the fully differential jet cross section}

We want to calculate the fully differential cross section of an $n$-jet
final state characterized by the jet-axis momenta $J_1,\ldots,J_n$ 
using the inclusive version of the jet algorithm specified in the previous
subsection. The jet event kinematics are given by
\bea\label{JetKin}
x_a\,P_a+x_b\,P_b &=& J_1+\cdots+J_n\ ,\nn
            J_i^2 &=& 0\ ,
\eea
where $P_{a,b}$ denote the incoming hadron momenta and
\beq
x_{a,b}\,P_{a,b}\;=\;x_{a,b}\,\frac{\sqrt{S}}{2}\,(1,\pm 1,0,0)\ .
\eeq
The collider energy is given by $\sqrt{S}$ and the momentum fractions
$x_a$ and $x_b$ are calculated from the reconstructed jets:
\beq
x_{a,b}\;=\;\frac{1}{\sqrt{S}}
          \;\sum_{i=1}^n\,p_T^{(i)}\,e^{\pm y_i}\ ,
\eeq
using the transverse momenta and rapidities of the jets,
\be
p_T^{(i)}\;=\;\sqrt{(p_x^{(i)})^2+(p_y^{(i)})^2}
\qquad\quad\mbox{and}\qquad\quad
y_i\;=\;\frac{1}{2}\,\log\left(\frac{E^{(i)}+p_z^{(i)}}{E^{(i)}-p_z^{(i)}}\right)\ ,
\ee
respectively. Note that here we have used the convention
$p=(E,p_z,p_x,p_y)$.

We define the differential cross section of the jet observable
${\cal O}$\/ as
\beq\label{eq:O}
\frac{d\,\sigma}{d\,{\cal O}}\;=\;\frac{1}{n!}\,\!
\int d\,\Phi\big(x_aP_a,x_bP_b\mapsto J_1,\ldots,J_n\big)\
d\,x_a\,d\,x_b\
\delta\Big({\cal O}-{\cal O}\big(J_1,\ldots,J_n\big)\Big)\
\frac{d^{(n)}\sigma}{d\,J_1\cdots d\,J_n}\ ,
\eeq
where the jet phase space is given by
\beqa
d\,\Phi\big(x_aP_a, x_bP_b\mapsto J_1,\ldots,J_n\big)
&=&
\left(\prod_{i=1}^n d^{(4)}J_i\ \delta(J_i^2)\ \theta(E_i)\right)
\delta(x_a P_a+x_b P_b-J_1-\cdots-J_n)\nnb\\[3mm]
&=&
\left(\prod_{i=1}^n\frac{d^{(3)}\vec{J_i}}{2\,E_i}\right)
\delta(x_a P_a+x_b P_b-J_1-\cdots-J_n)\ .
\eeqa
We consider all jets as indistinguishable and hence have to
introduce the ``identical-jets'' averaging factor of $(n!)^{-1}$.
The fully differential jet cross section at LO is given by
\be\label{eq:LO}
\frac{d^{(n)}\sigma_\trm{\tiny LO}}{d\,J_1\cdots d\,J_n}\;=\;
\frac{(2\pi)^{4-3n}}{2\,x_ax_b\,S}\
\sum_{\{f_af_b\to f_1\cdots f_n\}}F_{f_a}(x_a)\,F_{f_b}(x_b)\
\left|\overline{\cal M}^{(0)}\big(x_aP_a,x_bP_b;J_1,\ldots,J_n\big)\right|^2\ ,
\ee
where the $F_{f_{a,b}}$ are the parton density functions of the
partons in the beam particles and $\big|\,\overline{\cal M}^{(0)}\big|^2$
is the squared LO scattering amplitude, spin/color summed (averaged)
over final (initial) states. The flavor sum runs over all possible,
distinguishable partonic subprocesses $f_af_b\to f_1\cdots f_n$ that
contribute to the jet final state.\footnote{The flavor labels $f_i$
  denote gluons and massless (anti-)quarks. We omit specifying other
  than the partonic flavors for reasons of keeping the notation
  simple. For example, we could have a vector boson decaying
  leptonically in all subprocesses.}
For example, a $q\bar q+(n-2)g$\/ final state has $n(n-1)$ distinct
flavor terms to be added for one specific initial-state configuration.
This way we account for all ways of assigning the distinguishable
partons of the final state $f_1,\ldots,f_n$ to the jets $J_1,\ldots,J_n$.
Note that at LO no phase-space integration is left for the fully
differential jet cross section. However, using our definition, the
fully differential jet cross section will be symmetric under any
exchange of partons without the need of integrating over phase space.
Once one does the phase-space integration, as in \Eq{eq:O}, the usual
symmetry factors are recovered.

Because the jet algorithm preserves explicit momentum conservation and
keeps the jets massless, we can define a $K$-factor per jet
phase-space point $J_1,\ldots,J_n$. The NLO corrections to the fully
differential jet cross section can hence be written as
\beq\label{eq:K}
\frac{d^{(n)}\sigma_\trm{\tiny NLO}}{d\,J_1\cdots d\,J_n}\;=\;
K_\trm{\tiny NLO}\big(J_1,\ldots,J_n\big)\;\times\;
\frac{d^{(n)}\sigma_\trm{\tiny LO}}{d\,J_1\cdots d\,J_n}\ .
\eeq
In the remainder of the paper we will derive the expression for
$K_\trm{\tiny NLO}$ and develop, as a proof-of-principle, a Monte
Carlo integrator for the explicit evaluation of the $K$-factor for the
pure gluonic contribution of an $n$-jet event at a hadron collider.
The $K$-factor is composed of three contributions, the Born
contribution expressed as ``1'', the virtual contribution, $V$, and
the bremsstrahlung part, $R$. We write
\beqa\label{KfactorDef}
K_\trm{\tiny NLO}\big(J_1,\ldots,J_n\big)
&=&
1\;+\;\widetilde{V}\big(J_1,\ldots,J_n\big)\;+\;
\widetilde{R}\big(J_1,\ldots,J_n\big)
\nnb\\[3mm]&=&
1\;+\;\left(\frac{\alpha_S N_\mrm{C}}{2\pi}\right)
\Big(V(J_1,\ldots,J_n)\;+\;R(J_1,\ldots,J_n)\Big)\ ,
\eeqa
where we have factorized the strong-coupling expansion parameter
$\alpha_S/2\pi$ and the color factor $N_\mrm{C}$. \Eq{KfactorDef}
expresses the cancellation of infra-red singularities per jet
phase-space point, $\widetilde{V}$\/ and $\widetilde{R}$\/ on their
own diverge but their sum gives a finite contribution to the
$K$-factor. For the calculation of the virtual corrections, many
packages have been developed~\cite{Berger:2008sj,Giele:2008bc,
Ellis:2008qc,Lazopoulos:2008ex,Ellis:2009zw,Winter:2009kd,Berger:2009zg,
vanHameren:2009dr,Melnikov:2009dn,Bevilacqua:2009zn,Giele:2009ui,
Bevilacqua:2010ve,Mastrolia:2010nb,Badger:2010nx,Hirschi:2011pa},
which can be readily used to calculate this part of the $K$-factor. On
the other hand, the calculation of the bremsstrahlung contribution
$R$\/ requires a careful derivation.

To summarize, for the calculation of a jet observable, we generate,
using \Eq{eq:O}, the jet configurations contributing to the specific
value of the observable. For each generated jet phase-space point, we
calculate the LO weight according to \Eq{eq:LO} and the NLO
re-weighting multiplicative $K$-factor as given in \Eq{eq:K}.


\section{The Forward Branching Phase-Space Generator}
\label{sec:FBPS}

The explicit construction of the FBPS generator proceeds in several
steps. In \Sec{subsec:sector} the first step is taken by the
decomposition of the bremsstrahlung phase space into sectors using the
event resolution function given by the jet algorithm. Each sector is
defined through the jet algorithm selecting an unique triplet of
partons to be clustered. Next, owing to the invertibility of the
clustering, we develop in \Sec{subsec:forward} the real-emission
phase-space formalism based on forward branching off the Born level
jet configurations. In \Sec{subsec:brancher}, we derive the specific
forward branchers, which fill each sector such that the $3\to2$
cluster map given by the jet algorithm will recombine the three
partons to the same two jets specified by the jet phase-space point.
Finally, in \Sec{subsec:PS-numerics}, the procedure is validated using
the \textsc{Rambo} flat phase-space generator.

\subsection{An invertible sector decomposition of phase space}
\label{subsec:sector}

The bremsstrahlung contribution to the jet cross section, with the jet
kinematics specified in \Eq{JetKin}, is given by
\beqa\label{eq:basic-brem}
\lefteqn{
  \widetilde{R}\big(J_1,\ldots,J_n\big)\;=\;\left(
  \frac{d^{(n)}\sigma_\trm{\tiny LO}}
       {d\,J_1\cdots d\,J_n}\big(J_1,\ldots,J_n\big)\right)^{-1}
       \;\times\ \frac{(2\pi)^{1-3n}}{2\,S}\ \times\ \frac{1}{(n+1)!}}
\nnb\\[3mm]&\times&
\sum_{\{f_af_b\to f_1\cdots f_n f_{n+1}\}}\
\int d\,\Phi\big(\hat x_aP_a,\hat x_bP_b\mapsto p_1,\ldots,p_{n+1}\big)\
 d\,\hat x_a\ d\,\hat x_b\
\frac{F_{f_a}(\hat x_a)\,F_{f_b}(\hat x_b)}{\hat x_a\hat x_b}
\nnb\\[3mm]&\times&
\Delta_\mrm{jet}\big(J_1,\ldots,J_n\mid p_a,p_b,p_1,\ldots,p_{n+1}\big)\
\left|\overline{\cal M}^{(0)}\big(p_a,p_b;p_1,\ldots,p_{n+1}\big)\right|^2\ ,
\eeqa
where $p_a=\hat x_aP_a$ and $p_b=\hat x_bP_b$. Compared to the LO
case, we now have to sum over all partonic subprocesses with one more
parton in the final state. As before, cf.\
\Eqs{eq:O}~and~(\ref{eq:LO}), the flavor sum in combination with the
(identical-particle) averaging factor $1/(n+1)!$ guarantee the correct
symmetry factors for the $(n+1)$-particle final states.

The generalized jet delta-function $\Delta_\mrm{jet}$ decides whether
a specific subprocess with its parton kinematics contributes to the
bremsstrahlung factor $\widetilde{R}$ at the jet phase-space point
$J_1,\ldots,J_n$. This jet delta-function is equal to unity if the jet
algorithm clusters the parton momenta $p_a,p_b,p_1,\ldots,p_{n+1}$ to
the jet-axis four-vectors $J_1,\ldots,J_n$ and is zero otherwise. The
function (as used here) is flavor blind -- the jets as well as the
partons are indistinguishable, therefore, $\Delta_\mrm{jet}$ has to be
symmetric under any exchange of jet and parton momenta. To integrate
over the jet delta-function, we make use of the $3\to2$ clustering
algorithm discussed in the previous section. This allows us to expand
the jet delta-function over a sum of dipoles, each selecting three
partons, which will be clustered by the jet algorithm to two jets and,
as a result, a value for the resolution parameter $R_{ij;k}$ will be
returned for any of these combinations. The expansion can then be
written as follows:
%
%
\beqa\label{eq:jet-delta}
\lefteqn{
  \Delta_\mrm{jet}\big(J_1,\ldots,J_n\mid p_a,p_b,p_1,\ldots,p_{n+1}\big)}
\nnb\\[7mm]&=&
\delta(x_a-\hat{x}_a)\,\delta(x_b-\hat{x}_b)\ \times
\nnb\\[2mm]&&
\sum_{\genfrac{}{}{0pt}{1}{i\neq\,j}{i,j=1,\ldots,n}}\ \sum_{k\,\in\,S_{n+1}}
\Delta_\mrm{jet}\big(J_i;J_j\mid p_{k_i},p_{k_{n+1}};p_{k_j}\big)\ \
\frac{\Theta^\mrm{veto}_{k_ik_{n+1};k_j}}{2}\
\prod_{\genfrac{}{}{0pt}{1}{s\neq\,i,j}{s=1,\ldots,n}}\delta(J_s-p_{k_s})
\nnb\\[5mm]&+&\delta(x_b-\hat{x}_b)
\sum_{\genfrac{}{}{0pt}{1}{k\,\in\,S_{n+1}}{i=1,\ldots,n}}
\Delta_\mrm{jet}\big(J_i\mid p_a;p_{k_{n+1}},p_{k_i}\big)\ \
\Theta^\mrm{veto}_{a k_{n+1};k_i}\
\prod_{\genfrac{}{}{0pt}{1}{s\neq\,i}{s=1,\ldots,n}}\delta(J_s-p_{k_s})
\nnb\\[5mm]&+&\delta(x_a-\hat{x}_a)
\sum_{\genfrac{}{}{0pt}{1}{k\,\in\,S_{n+1}}{i=1,\ldots,n}}
\Delta_\mrm{jet}\big(J_i\mid p_b;p_{k_{n+1}},p_{k_i}\big)\ \
\Theta^\mrm{veto}_{b k_{n+1};k_i}\
\prod_{\genfrac{}{}{0pt}{1}{s\neq\,i}{s=1,\ldots,n}}\delta(J_s-p_{k_s})\ ,
\eeqa
where the sums are over all possible pairs of jet-axis momenta
$J_1,\ldots,J_n$ and permutations of the brems\-strah\-lung
four-momenta $p_1,\ldots,p_{n+1}$. The index vector $k$\/ describes
the elements of the permutations $S_{n+1}$ of the set
$\{1,2,\ldots,n+1\}$. The permutation sum ensures that all
dipole--spectator configurations {\em and}\/ their respective $(n-2)!$
phase-space combinations in the leftover parton momenta are taken into
account.

The sector veto (or, just as well, jet resolution) cut
$\Theta^\mrm{veto}_{ir;j}$ implements the first two steps of our jet
algorithm proposed in \Sec{subsec:algodef}.
According to \Eq{eq:step1}, we evaluate
\be
R_\mrm{min}\;\equiv\;R_{\bar v\bar w}\;\equiv\;\min_{vw}R_{vw}
\ee
where $v,w=1,\ldots,n+1$ and $v<w$. If we take $\bar v,\bar w\neq
r=1,\ldots,n+1$ and define
\be
R'_\mrm{min}\;\equiv\;\min_r R_{\bar v\bar w;r}\;=\;
\min_r\ \min(R_{\bar vr},R_{\bar wr})\ ,
\ee
cf.\ \Eq{eq:step2}, we can formulate this cut as
\be\label{sectorcut}
\Theta^\mrm{veto}_{im;j}\;=\;
\theta\left(R_\mrm{min}-R_{im}\right)\
\theta\left(R^\prime_\mrm{min}-R_{im;j}\right)\ ,
\ee
using the Heaviside step function $\theta(x)$, which equals one if
$x\ge0$ and is zero otherwise.\footnote{The jet clustering of
  \Sec{subsec:algodef} is symmetric under the exchange of the emitter
  and emitted parton ($R_{wv}=R_{vw}$ and $R_{wv;r}=R_{vw;r}$).
  Therefore, we do not have to consider $w>v$\/ to determine the
  $R_\mrm{min}$ and $R^\prime_\mrm{min}$.}
Step 3 and 4 of the proposed jet algorithm are executed when one
computes whether the generic $\Delta_\mrm{jet}$ building blocks are
different from zero. They will return one, only if the parton momenta
reconstruct to the two given jet momenta or, in the initial--final
state cases, to the one given jet momentum. More specifically, we
understand $\Delta_\mrm{jet}(J;J'\mid p_i,p_m;p_j)\equiv1$, if and
only if $J\equiv p_{im}+(1-\gamma)\,p_j$ and $J'\equiv\gamma\,p_j$,
cf.\ \Eq{eq:step3}. Neither in the vice versa case $J\leftrightarrow J'$,
nor in any other combination we find the generic $\Delta_\mrm{jet}\neq0$.
This way we avoid double counting when we permute over all parton
momentum configurations. The emitter--emitted parton symmetry in
$\Theta^\mrm{veto}_{im;j}$, however, leads to double counting the same
event in the final--final parton sum of \Eq{eq:jet-delta}, which we
remove by multiplying the factor $1/2$ to the veto.\footnote{Later on,
  we resolve this issue by partitioning the phase space further
  according to the different parton emitter settings.} Since the order
of the bremsstrahlung momenta is permuted, we guarantee that all
combinations are tested (in $\Delta_\mrm{jet}$ as well as the product
of the $\delta(J_s-p_{k_s})$ terms) to make sure that the
bremsstrahlung events are selected, which match the considered jet
phase-space point kinematics and, therefore, give a contribution to
$\widetilde{R}(J_1,\ldots,J_n)$.

We observe in \Eq{eq:jet-delta} that the jet delta-function breaks up
phase space in two types of sectors: final--final state sectors and
initial--final state sectors. Note that in principle there could be an
initial--initial state sector as well. However, this will only occur
if we allow for the build-up of beam jets.

We finally note that for each jet phase-space point exactly one sector
contributes. With $R'_\mrm{min}$ (based on $R_\mrm{min}$) a global
event resolution measure is given, which depends on all initial-state
and jet momenta. This partitioning of phase space is dictated by the
event resolution function given by the jet algorithm, see
\Eq{eq:step1}. As a consequence the phase space is invertible: given
the jet four-momenta $J_i$ and $J_j$ one can -- by inverting the
cluster map of the jet algorithm -- generate the three-parton
configurations for each sector, which will cluster back to these two
initial jets. Such a forward branching Monte Carlo integrator exactly
integrates out the internal jet structure. In the next section we will
formulate these forward branchers.

\subsection{Phase-space construction through forward branching}
\label{subsec:forward}

To construct the forward brancher for a sector, we have to integrate
\Eq{eq:basic-brem} over the jet delta-function of \Eq{eq:jet-delta}.
The jet delta-function selects those $(n+1)$-parton final states, 
which reconstruct to the given $n$-jet
phase-space point. We can turn the approach around and use 
the jet delta-function as a
prescription to explicitly generate the $n+1$ bremsstrahlung parton
momenta given the $n$-jet momenta. 
This establishes the forward-branching picture, which in
addition allows for the avoidance of the dipole and permutation sums
of \Eq{eq:jet-delta}. To see this, we can write down the final--final
state piece of \Eq{eq:basic-brem} for a single subprocess neglecting
all prefactors, including the Born matrix element:
\bea
\lefteqn{
  \widetilde{R}_\trm{\tiny FF}\big(J_1,\ldots,J_n\big)}
\nnb\\[4mm]&\sim&
\sum_{\genfrac{}{}{0pt}{1}{i\neq\,j}{i,j=1,\ldots,n}}\ \sum_{k\,\in\,S_{n+1}}
d\,\Phi\big(J_{ij}\mapsto p_{k_i},p_{k_{n+1}},p_{k_j}\big)\ \
\Delta_\mrm{jet}\big(J_i;J_j\mid p_{k_i},p_{k_{n+1}};p_{k_j}\big)
\nnb\\[4mm]&\times&
\frac{\Theta^\mrm{veto}_{k_ik_{n+1};k_j}}{2}\ \ \times\ \
\left|\overline{\cal M}^{(0)}\big(x_aP_a,x_bP_b;
{\{p_{k_s}=J_s\}}_{s\neq i,j},p_{k_i},p_{k_{n+1}},p_{k_j}\big)\right|^2
\eea
where $J_{ij}=J_i+J_j$. We observe that the phase-space integration
has been broken up into many factorized pieces of splitting dipoles.
Furthermore, we can exchange the order in performing the dipole
phase-space integrations and summations. The integration over the
three-parameter phase spaces can be accomplished through Monte Carlo
techniques. We can treat the explicit dipole and permutation sums
similarly: instead of carrying them out, we can choose dipole and
parton configurations at random.\footnote{Owing to the flavor
  blindness of the jet definition, the parton configurations have to
  be varied too as long as the chosen subprocess contains different
  parton flavors. We can use any set of indistinguishable particles
  though to reduce the initial number of $(n+1)!$ possibilities.}
We just have to keep track of and include possible weights that may
occur in the selection of dipoles and bremsstrahlung partons.

\Eqs{eq:LO}~and~(\ref{eq:basic-brem}) have complete flavor sums
running over all possible subprocesses that contribute to the $n$-jet
and $(n+1)$-jet final states, respectively. We can maintain this
inclusive-flavor approach in the forward generation of the
real-emission events. No knowledge of the
particular LO process and its flavors is needed apart from the given
set of the jet-axis momenta, which we interpret as the initial
four-vectors before the parton branching. The forward branching
occurs, in principle, independently of flavor; the pure generation of
the bremsstrahlung momenta in fact has no flavor dependence. The only
place where flavor conditions enter is in combining phase space with
the matrix element for the randomly chosen subprocess containing
$2\to n+1$ strongly interacting particles: the number of $3\to2$
clusterings as given by combinatorics may reduce owing to flavor
constraints.\footnote{As an alternative the FBPS generator may be
  designed such that the parton flavors are treated as in dipole
  showers. For example, select
  a flavor assignment at LO as in \Eq{eq:LO}, now consider an
  initial-state branching; if the LO subprocess has an incoming quark,
  there are two bremsstrahlung contributions: $q\to qg$\/ and $g\to
  q\bar q$\/ where the gluon and anti-quark are radiated off,
  respectively. The corresponding $n+1$ matrix element then determines
  the weight of the selected option.}

To simplify the discussion, we focus on the pure gluonic case.
Consequently, the flavor sums in \Eqs{eq:LO}~and~(\ref{eq:basic-brem})
collapse to single terms. We also can simply arrange to set the first
$n$\/ partonic four-vectors according to the jet-axis momenta. In the
final--final case for example, two of these, the emitter and spectator
momenta, will change owing to the generation of the additionally
emitted parton, which we can always choose to label by $n+1$. For
ordered amplitudes, one may insert the new parton right after the
emitter parton $l-1$ and shift all subsequent ones by one, $l\to l+1$.
Because of the forward construction of the parton momenta, the criteria
underlying the generic $\Delta_\mrm{jet}(J_i;J_j\mid p_i,p_r;p_j)$
terms will be satisfied by construction. Thus, these terms are
redundant and the constrained generation of $p_i$, $p_r$ and $p_j$
already accounts for the $(n+1)\,n\,(n-1)$ combinations of arranging
three partons to be clustered to the two jets picked for forward
branching. Still, with respect to the non-branching part of the final
state, we have $(n-2)!$ possibilities to assign the parton momenta
with certain jet momenta. However, owing to the symmetry of the final
state, the leftover permutation sum can be replaced by a
multiplicative factor. The $\Theta^\mrm{veto}_{ir;j}$ is included to
the three-parton phase space and acts as a phase-space cut
implementing the jet resolution criteria. Applying similar arguments
to the initial--final state cases, we can take all modifications and
write the result of the phase-space integration as
\beqa\label{RDef}
\lefteqn{
  \widetilde{R}\big(J_1,\ldots,J_n\big)\;=\;
  \frac{1}{(2\pi)^3}\ \frac{1}{n\,(n+1)}\ \left|
  \overline{\cal M}^{(0)}\big(J_a=x_aP_a,J_b=x_bP_b;J_1,\ldots,J_n\big)
  \right|^{-2}\;\times}
\nnb\\[7mm]&\Biggl[&
\sum_{\genfrac{}{}{0pt}{1}{i\neq\,j}{i,j=1,\ldots,n}}\frac{1}{n-1}\
\int d\,\Phi_\mrm{veto}\left(J_i,J_j\mapsto p_i,p_r,p_j\right)\
\left|\overline{\cal M}^{(0)}
\big(J_a,J_b;{\{p_s=J_s\}}_{s\neq i,j},p_i,p_r,p_j\big)\right|^2
\nnb\\[5mm]&+&
\sum^n_{j=1}\
\int d\,\Phi_\mrm{veto}\left(J_a,J_j\mapsto p_a,p_r,p_j\right)\
\frac{x_a\,F_{f_a}(\hat{x}_a)}{\hat{x}_a\,F_{f_a}(x_a)}\
\left|\overline{\cal M}^{(0)}
\big(p_a,J_b;{\{p_s=J_s\}}_{s\neq j},p_r,p_j\big)\right|^2
\nnb\\[5mm]&+&
\sum^n_{j=1}\
\int d\,\Phi_\mrm{veto}\left(J_b,J_j\mapsto p_b,p_r,p_j\right)\
\frac{x_b\,F_{f_b}(\hat{x}_b)}{\hat{x}_b\,F_{f_b}(x_b)}\
\left|\overline{\cal M}^{(0)}
\big(J_a,p_b;{\{p_s=J_s\}}_{s\neq j},p_r,p_j\big)\right|^2\quad\Biggr]\ .
\nnb\\[3mm]&&
\eeqa
Note that the $(n+1)!$ term in the denominator has been combined
with the multiplicative numerator factors $(n-2)!$ and $(n-1)!$ for
final--final and initial--final state branchings, respectively. Also,
$r\equiv n+1$ and $p_{a,b}=\hat x_{a,b}P_{a,b}$ while
$J_{a,b}=x_{a,b}P_{a,b}$. As before, the dipole
sum and the three-parameter phase space can be calculated using a
Monte Carlo integration, \ie $\sum\int\to(1/N_\trm{\tiny MC})
\sum_{i=1,\ldots,N_\trm{\tiny MC}} d\,\widetilde{R}^{(i)}$.

The dipole factorization of phase space is obvious from the equation
above. The $2\to3$ differential phase-space volumes can be described
by dipole or antenna phase-space factors, which are also used in
shower algorithms. In our calculation the combination with the matrix
element then fully specifies the branching, which in showers is
achieved only approximately by the use of the splitting function. The
final--final state antenna phase space,
cf.~\cite{Giele:1991vf,Weinzierl:1999yf}, is given by
\beq\label{PSff}
d\,\Phi_\mrm{veto}\big(J_i,J_j\mapsto p_i,p_r,p_j\big)\;=\;
\frac{\Theta^\mrm{veto}_{ir;j}}{2}\ \
\frac{\pi}{2}\ \frac{1}{s_{irj}}\ d\,s_{ir}\,d\,s_{rj}\ \frac{d\,\phi}{2\pi}\ ,
\eeq
whereas the initial--final state antenna phase space,
cf.~\cite{Giele:1993dj,Daleo:2006xa} is expressed as
\beq\label{PSif}
d\,\Phi_\mrm{veto}\big(J_a=x_aP_a,J_j\mapsto\hat{x}_aP_a,p_r,p_j\big)\;=\;
\Theta^\mrm{veto}_{ar;j}\ \
\frac{1}{2\pi}\ \frac{d^{(3)}\vec p_r}{2\,E_r}
\left(\frac{P_a\cdot J_j}{P_a\cdot J_j-P_a\cdot p_r}\right)\ .
\eeq
In the latter case, $p_j$ is given by momentum conservation, \ie
$x_aP_a-J_j=\hat{x}_aP_a-p_r-p_j$. The new momentum fraction
$\hat{x}_a$ can be calculated using the condition $p_j^2=0$; we obtain
\be
\hat{x}_a\;=\;x_a+\frac{J_j\cdot p_r}{P_a\cdot J_j-P_a\cdot p_r}\ .
\ee
In both cases above, the phase-space factors have to be supplemented
by the corresponding sector veto (or jet resolution) cut. The cuts
guarantee the partitioning of the bremsstrahlung phase space such that
no overlapping regions can emerge. The actual sector phase-space
volume is then measured by the Monte Carlo integration by means of the
sector veto cut.

The formulation of the effective FBPS generator used in the remainder
of the paper is now simple. Given an $n$-jet event, we generate the
bremsstrahlung events in the following manner: with probability
$(n-1)/(n+1)$, select a pair of final-state jets randomly and perform
a final--final state branching. Else, select one of the two incoming
partons and one final-state jet and perform an initial--final state
branching. \App{appendix:counting} gives more explanations regarding
this selection. The bremsstrahlung event now has $n+1$ final-state
partons, which reconstruct back to the original jet configuration
using our specific jet algorithm. We repeat the procedure until a
sufficient number of bremsstrahlung events have been generated to
estimate the $K$-factor for this particular jet event. This in fact is
the execution of the Monte Carlo integration, whose uncertainty can be
controlled by the number of generated Monte Carlo events per jet
phase-space point.

\subsection{Forward branchers}
\label{subsec:brancher}

To completely assemble the forward-branching buildup of the
bremsstrahlung phase space, we still have to define the generated
three-parton final state in terms of the original jets and the dipole
phase-space integration variables. In doing so we have to respect the
constraint that the jet algorithm clusters the three generated partons
back to the two initiator jets. In the next two subsections, we will
formulate the branchers, explicitly designed for this task.

\subsubsection{The final--final state brancher}
\label{subsubsec:ff-brancher}

From the final--final state phase-space factor of \Eq{PSff} we extract
the phase-space factor for the FBPS generator. We write
\beq\label{PSffdetail}
d\,\Phi_\mrm{veto}\big(J_i,J_j\mapsto p_i,p_r,p_j\big)\;=\;
\frac{\pi}{2}\ s_{irj}\ d\,y_{ir}\,d\,y_{rj}\ \frac{d\,\phi}{2\pi}\
\theta\left(1-y_{ir}-y_{rj}\right)\ \theta\left(y_{ij}-y_{rj}\right)\
\Theta^\mrm{veto}_{ir;j}
\eeq
where the sector veto cut has been introduced in \Eq{sectorcut}.
Several comments are in order. We have defined $y_{kl}=s_{kl}/s_{irj}$
with $s_{kl}=(p_k+p_l)^2$ and $s_{irj}=(p_i+p_r+p_j)^2=(J_i+J_j)^2$.
Apart from the kinematic constraint, $y_{ir}+y_{rj}\le1$, we added an
additional constraint, $y_{rj}\le y_{ij}$, which divides each sector
into two (sub)sectors, breaking  the
emitter--emitted parton ($i\leftrightarrow r$) symmetry. As a result,
the factor $1/2$ formerly present in \Eq{PSff} is dropped here. The
additional constraint is needed to accommodate for the integration
over asymmetric functions in $p_i$ and $p_r$, \eg over ordered
amplitudes (or over quark--gluon states, which is important for later
applications). The distinction is particularly important when we
combine the sectors with ordered matrix elements as each sector has
its own singularity structure. It manifests the notion of $p_i$, $p_r$
and $p_j$ respectively being the emitter, emitted parton and spectator
in the phase-space branching.

The algorithmic description of the final--final state brancher is as
follows: starting from two final-state jets $J_i$ and $J_j$;
\begin{enumerate}
\item generate the integration variables $y_{ir}$, $y_{rj}$ and
  $\phi/(2\pi)$ on the interval $(0,1]$ and fulfill the constraints
  $y_{ir}+y_{rj}\le1$ and $y_{rj}\le y_{ij}$. Notice that
  the sector veto cut $\Theta^\mrm{veto}_{ir;j}$ guarantees
  $y_{ir}<y_{rj}$, cf.\ \Eq{sectorcut}.
\item rescale the four-momenta $J_i$ and $J_j$:
  \be
  \left\{\begin{array}{l}
  k_i\;=\;J_i+\gamma\,J_j\ ,\\[2mm]
  k_j\;=\;(1-\gamma)\,J_j\ ,
  \end{array}\right.
  \ee
  where $\gamma=y_{ir}$ such that we find $k_i^2=y_{ir}\,s_{irj}$.
\item determine the final-state partons $p_i$ and $p_r$ by invoking
  the phase-space decay of $k_i\to p_i+p_r$ with the on-shell
  condition $p_i^2=p_r^2=0$.
\item set $p_j=k_j$.
\item pass the event if and only if the sector decomposition cut
  $\Theta^\mrm{veto}_{ir;j}$ has been satisfied. Assign the weight
  $s_{irj}\,\pi/2$ to the event.\footnote{This may be supplemented by
  a possible weight from the generation of the integration variables.
  For example, one may rewrite $ds$\/ as $s\,d(\log s)$, which would
  generate an additional weight to be included.}
\end{enumerate}
Using this construction procedure, the final--final state clustering
of the jet algorithm maps the partonic set $\{p_i,p_r,p_j\}$ created
through this FBPS generator back onto the jet pair $\{J_i,J_j\}$. Note
that each of the forward branchers would generate the ``same''
one-particle phase space, if the sector veto cut
$\Theta^\mrm{veto}_{ir;j}$ was removed from \Eq{PSffdetail}. An
additional integration over the $n$-jet phase space would generate
the whole $(n+1)$-particle bremsstrahlung phase space.

\subsubsection{The initial--final state brancher}
\label{subsubsec:if-brancher}

From the initial--final state phase-space factor of \Eq{PSif} we
extract the corresponding phase-space factor for the FBPS generator:
\bea\lefteqn{
d\,\Phi_\mrm{veto}\big(x_aP_a,J_j\mapsto\hat{x}_aP_a,p_r,p_j\big)}
\nnb\\[4mm]&=&
\frac{1}{2\pi}\ \frac{d^{(3)}\vec p_r}{2\,E_r}
\left(\frac{P_a\cdot J_j}{P_a\cdot J_j-P_a\cdot p_r}\right)
\theta\left(|s_{rj}|-|s_{ar}|\right)\ \Theta^\mrm{veto}_{ar;j}\ .
\eea
Because the initial-state parton is distinct from the final-state
partons, no additional ordering requirements are present. Note that --
as in the previous case -- by removing the sector veto cut
$\Theta^\mrm{veto}_{ar;j}$, one allows for this forward brancher to
generate the whole $(n+1)$-parton bremsstrahlung phase space. The
algorithmic construction underlying this FBPS generator is outlined
below. It is set up such that the initial--final state clustering of
the jet algorithm maps the generated triplet of momenta
$\{\hat x_aP_a,p_r,p_j\}$ back onto the initial--final jet pair
$\{x_aP_a,J_j\}$. The generation of the momenta then proceeds as
follows: starting from an initial-state parton $x_aP_a$ and a
final-state jet $J_j$;
\begin{enumerate}
\item generate the one-particle phase-space momentum $\vec p_r$ within
  the appropriate integration boundaries.
\item having generated $p_r$, calculate
  \be
  \left\{\begin{array}{l}
  z\;=\;\hat x_a/x_a\;=\;
  1\;+\;(J_j\cdot p_r)\big/\big[x_aP_a\cdot(J_j-p_r)\big]\ ,\\[2mm]
  p_j\;=\;J_j-p_r+(z-1)\,x_aP_a\ .
  \end{array}\right.
  \ee
\item pass the event with weight
  $\frac{1}{2\pi}\left(\frac{P_a\cdot J_j}{P_a\cdot(J_j-p_r)}\right)$
  assigned, if and only if the jet resolution cut
  $\Theta^\mrm{veto}_{ar;j}$ has been satisfied.
\end{enumerate}

\subsection{Numerical validation of the phase space generator}
\label{subsec:PS-numerics}

\begin{figure}[p!]
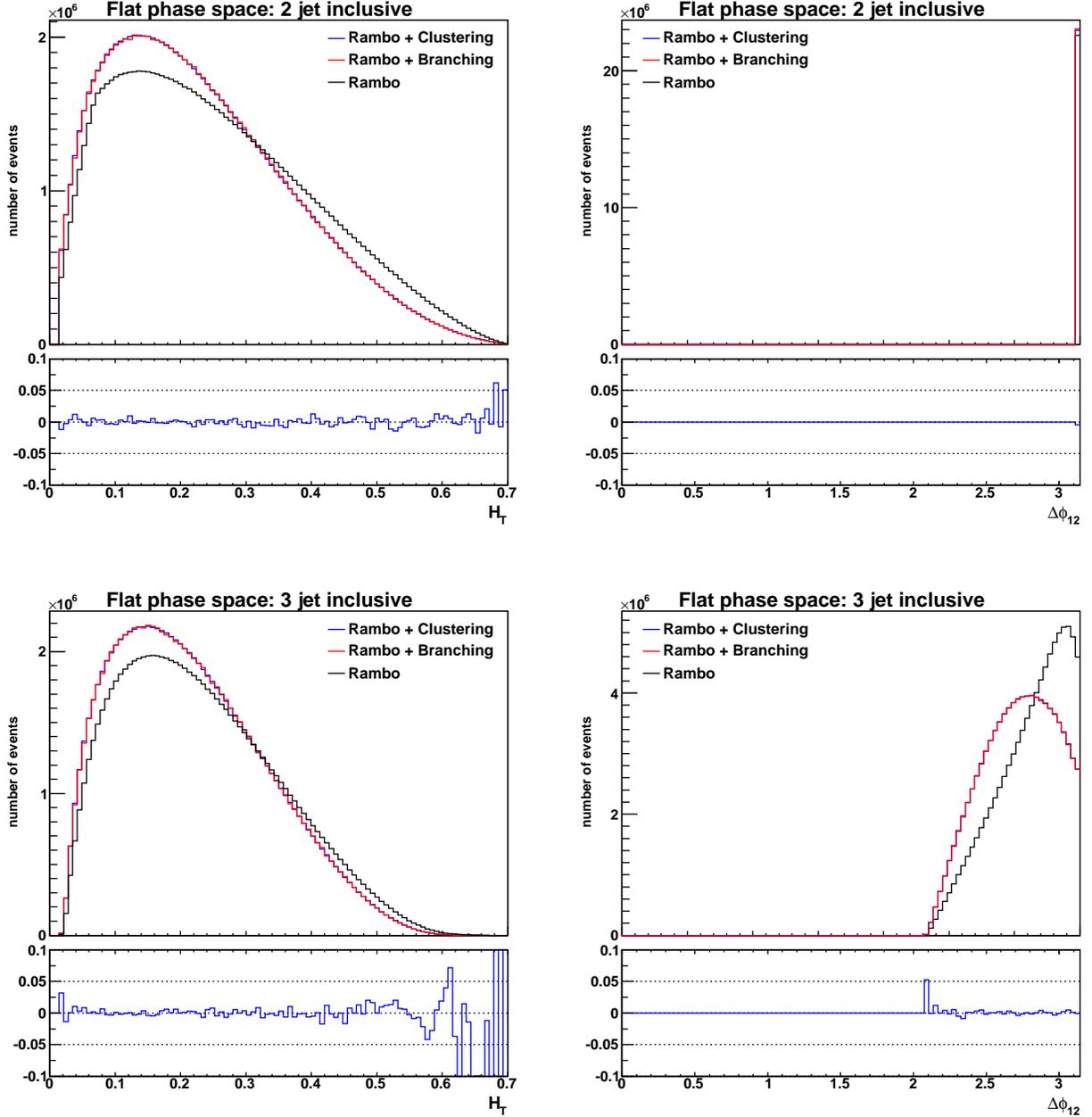

  \centerline{\includegraphics[clip,width=0.48\columnwidth,angle=-90]{%
      graphs/PS2_i1000_n100000_b10.epsi}}\vskip10mm
  \centerline{\includegraphics[clip,width=0.48\columnwidth,angle=-90]{%
      graphs/PS3_i1000_n100000_b10.epsi}}\vskip5mm
  \caption{\label{fig1}
    The $H_T$ and $\Delta\phi_{12}$ pure phase-space comparisons
    between clustered $(n+1)$-particle \textsc{Rambo} events (blue
    lines) and forward-branched $n$-particle \textsc{Rambo} events
    (red lines) for $n=2$-jet (upper graphs) and $n=3$-jet (lower
    graphs) configurations. The lower panels show the ratio between
    the clustered and branched predictions minus one. The black lines
    represent the results of the corresponding $n$-particle
    \textsc{Rambo} generations. To focus on the phase-space
    validation, the matrix-element and PDF weights have not been
    included in these calculations.}
\end{figure}

\begin{figure}[p!]
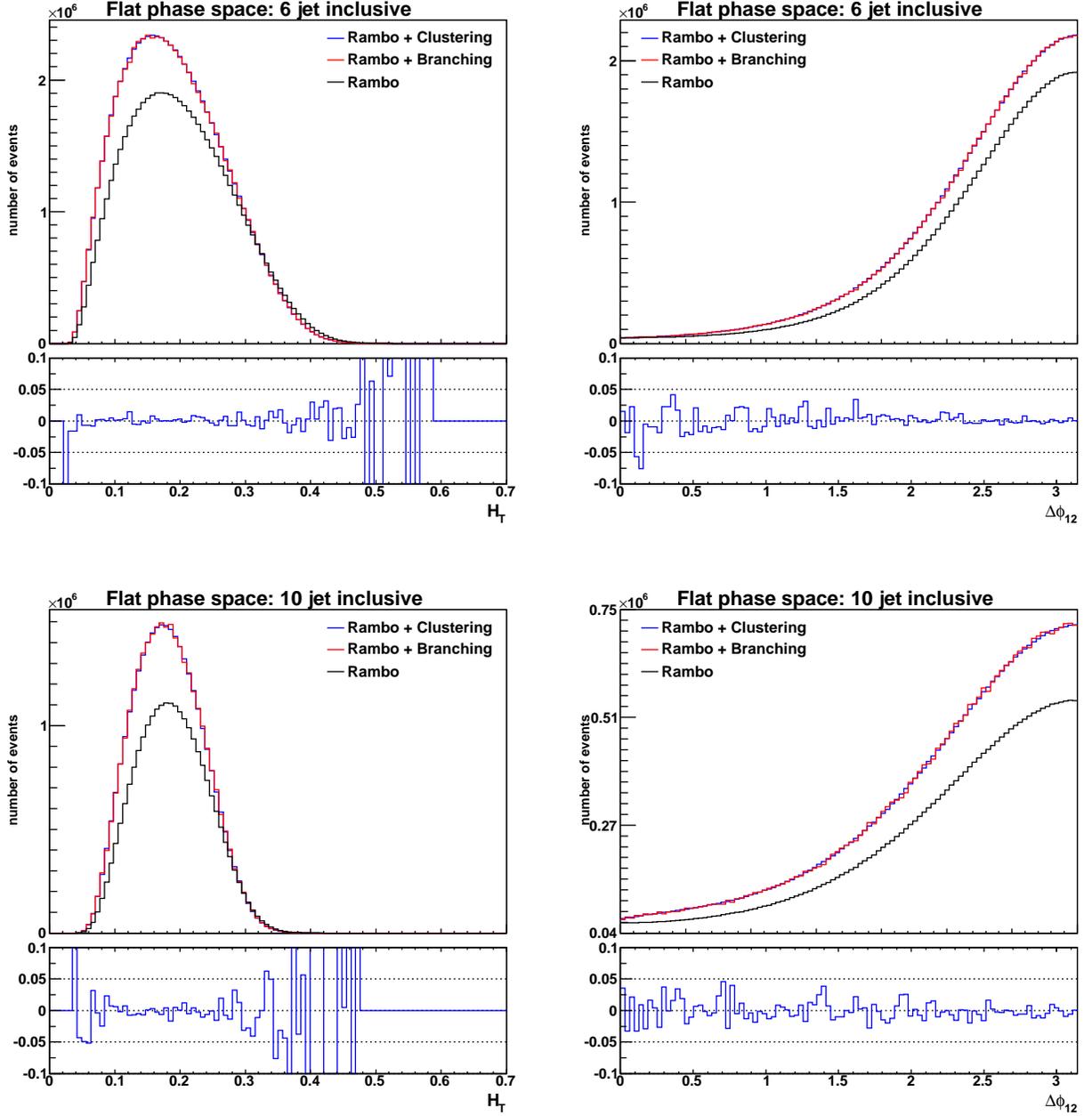

  \centerline{\includegraphics[clip,width=0.48\columnwidth,angle=-90]{%
      graphs/PS6_i1000_n100000_b10.epsi}}\vskip10mm
  \centerline{\includegraphics[clip,width=0.48\columnwidth,angle=-90]{%
      graphs/PS10_i1000_n100000_b10.epsi}}\vskip5mm
  \caption{\label{fig2}
    The $H_T$ and $\Delta\phi_{12}$ pure phase-space comparisons
    between clustered $(n+1)$-particle \textsc{Rambo} events (blue
    lines) and forward-branched $n$-particle \textsc{Rambo} events
    (red lines) for $n=6$-jet (upper graphs) and $n=10$-jet (lower
    graphs) configurations. The lower panels show the ratio between
    the clustered and branched predictions minus one. The black lines
    represent the results of the corresponding $n$-particle
    \textsc{Rambo} generations. To focus on the phase-space
    validation, the matrix-element and PDF weights have not been
    included in these calculations.}
\end{figure}

We want to verify the FBPS generator on itself,
before we use the generator to calculate the
$K$-factor for an $n$-jet phase-space point. 
For this purpose we do not
add in the contributions stemming from the matrix elements and
PDFs. This is an important validation to ensure 
the correct treatment of the weight generation during
the build-up of the bremsstrahlung phase space.
We use the flat phase-space generator
\textsc{Rambo}~\cite{Kleiss:1985gy} for the numerical validation of
the FBPS generator. The phase space generated by Rambo,
$d\,{\cal R}_n$, is connected to the customary flat phase space
$d\,\Phi$ through:
\bea\label{PSRambo}
\lefteqn{
d\,\Phi\big(x_aP_a,x_bP_b\mapsto J_1,\ldots,J_n\big)\;=\;}
\nnb\\[4mm]&&
\left(\frac{\pi}{2}\right)^{(n-1)}
\left(\frac{(x_ax_b\,S)^{(n-2)}}{(n-1)!\,(n-2)!}\right)\
d\,{\cal R}_n\big(x_aP_a,x_bP_b\mapsto J_1,\ldots,J_n\big)\ .
\eea
For this test, we do not include parton density functions; 
instead we choose the parton fractions uniformly between zero
and one. We define an $n$-jet event at collider energy of 7~TeV using
the following selection criteria for the $n$\/ jets regarding their
transverse momentum, rapidity and geometrical jet--jet separation:
$p^{(i)}_T>250$~GeV, $|\,\!y_i|<2.0$ and $\Delta R_{ij}>0.5$. We
present the results of our tests in \Figs{fig1}~and~\ref{fig2} where
we exemplify the FBPS validation by means of comparisons of
distributions for two distinct observables. Here, we define
\be
H_T\;=\;\sqrt{\frac{\sum\limits_i \left|\vec p^{\,(i)}_T\right|^2}{S}}\ ,
\ee
as a dimensionless (scaled) variant of an ordinary $H_T$ variable
where instead of using the scalar sum of the jet $p_T$s, the squared
quantities have been summed up. With the second observable, we want to
look into an angular distribution, namely the azimuthal angle between
the two leading jets. This variable is calculated by
\be
\Delta\phi_{12}\;=\;\arccos{\left(
  \frac{\;\;\vec p^{\,(1)}_T\cdot\vec p^{\,(2)}_T\;}
       {\;|\vec p^{\,(1)}_T|\,|\vec p^{\,(2)}_T|\;}\right)}\ .
\ee
For various $n$-jet final states, the $H_T$ and $\Delta\phi_{12}$
distributions resulting from our phase-space tests are displayed in
\Figs{fig1}~and~\ref{fig2}. All plots contain three curves labeled by
``Rambo'', ``Rambo+Branching'' and ``Rambo+Clustering'' (represented
by the black, red and blue lines, respectively); in the corresponding
lower panes, we always show the ratio, subtracted by one, between the
latter two predictions. In general we observe steeper tails in the
$H_T$ spectra for an increasing number of jets. The $\Delta\phi_{12}$
distributions show the opposite behavior; their low angle-difference
bins become more populated owing to a larger amount of phase space
being filled. For the same reason of enhanced phase-space filling, the
total energy is shared among more jets, which leads to the suppression
of the $H_T$ tails.

We now explain the three different predictions, which we use for the
validation and show in the figures. The ``Rambo'' curves are obtained
from jet momenta generated according to \Eq{PSRambo} where we define
the ``LO'' $n$-jet phase space as the $n$-particle uniform phase
space. The jet momenta satisfy the acceptance cuts given above. To
produce the ``NLO'' $n$-jet phase space, we generate $n+1$ particles
in flat phase space with the help of \textsc{Rambo} and apply our jet
algorithm to find $n$\/ jets from which we can calculate the jet
observables. Again, these jets have to fulfill the acceptance
criteria. This procedure gives us the ``Rambo+Clustering'' predictions
in \Figs{fig1}~and~\ref{fig2}. The differences seen between ``Rambo''
and ``Rambo+Clustering'' visualize the pure phase-space effect when
generating the ``NLO'' corrections.

We can now validate the FBPS generator: we construct ``LO'' $n$-jet
configurations using the flat phase-space generator as we did for the
``Rambo'' predictions. For each configuration, we subsequently
generate bremsstrahlung events using the FBPS. As the generated events
always reconstruct back to the originating $n$-jet event, we only have
to average over the generated event weights.\footnote{Of course, we
  also cross-checked that the backward clustering indeed recovered the
  $n$-jet phase-space point.}
This determines the ``Rambo+Branching'' curves, which have to coincide
-- apart from statistical fluctuations -- with the respective curves
of the ``Rambo+Clustering'' procedure. The ratio plots in
\Figs{fig1}~and~\ref{fig2} illustrate how well this is achieved by
directly comparing the flat phase-space ``NLO'' predictions and the FBPS
generated ``NLO'' predictions. We see excellent agreement between the
two results and thereby validate the FBPS generator. It is interesting
to note that at ``NLO'' the $\Delta\phi_{12}$ distribution for two
jets does not show the usual feature of de-correlating. Because the
jet algorithm is infra-red safe for all jet observables, the two jets
are always exactly back-to-back (in the azimuthal angle) for both
``LO'' and ``NLO'', \ie this dijet observable is not affected by the
initial-state radiation.


\section{The Gluonic Jet Generator}
\label{sec:POP}

As a proof-of-principle we describe in this section a generator, which
calculates the NLO $K$-factor per jet phase-space point for the pure
gluonic part of $n$-jet production at hadron colliders in the
leading-color approximation.

The determination of the $K$-factor for a given $n$-jet event requires
the evaluation of a single virtual event and a three-dimensional Monte
Carlo integration over the bremsstrahlung phase-space sectors defined
by the jet resolution scale. This is discussed in \Sec{subsec:real}.
Finally, in \Sec{subsec:jet-numerics} our results are presented: we
verify the cancellation of the slicing parameter in the $K$-factor and
the evaluation of the full leading-color NLO corrections up to
$15$-jet configurations.

\boldmath
\subsection{Evaluation of the $K$-factor}\unboldmath
\label{subsec:real}

In terms of ordered amplitudes $m$, the fully differential gluonic
leading-order, leading-color contribution to the $n$-jet cross section
is given as
\beqa\label{eq:orderedLO}
\frac{d^{(n)}\sigma_\trm{\tiny LO}}{d\,J_1\cdots d\,J_n} &=&
\frac{(2\pi)^{4-3n}}{2\,x_ax_b\,S}\
F_{g}(x_a)\,F_{g}(x_b)\ N_\mrm{C}^{n}\left(N^2_\mrm{C}-1\right)
\nnb\\[4mm]&\times&
\left[\,\sum_{\sigma\,\in\,S_{n+1}}
\left|\overline{m}^{(0)}\big(x_aP_a,
J_{\sigma_1},J_{\sigma_2},\ldots,J_{\sigma_{n+1}}\big)\right|^2
\;+\;{\cal O}\left(\frac{1}{N_\mrm{C}^2}\right)\,\right]\ ,
\eeqa
where the color, $N_\mrm{C}$, dependence has been made explicit. The
sum is over all $(n+1)!$ different permutations of the ordered
amplitudes and the $|\overline{m}^{(0)}|^2$ are the corresponding
helicity averaged/summed squared matrix elements. Note that the
$\sigma_i$ take values from $b,1,\ldots,n$ where $J_b=x_bP_b$. We can
now define the fully differential NLO cross section through ordered
$k$-factors,
\beqa\label{eq:orderedNLO}\lefteqn{
\frac{d^{(n)}\sigma_\trm{\tiny NLO}}{d\,J_1\cdots d\,J_n}\;\,=\;\,
\frac{(2\pi)^{4-3n}}{2\,x_ax_b\,S}\
F_{g}(x_a)\,F_{g}(x_b)\ N_\mrm{C}^{n}\left(N^2_\mrm{C}-1\right)\;\,\times}
\nnb\\[4mm]&&
\left[\,\sum_{\sigma\,\in\,S_{n+1}}
\overline{k}_\trm{\tiny NLO}\big(J_a,
J_{\sigma_1},J_{\sigma_2},\ldots,J_{\sigma_{n+1}}\big)
\left|\overline{m}^{(0)}\big(J_a,
J_{\sigma_1},J_{\sigma_2},\ldots,J_{\sigma_{n+1}}\big)\right|^2
\,+\;{\cal O}\left(\frac{1}{N_\mrm{C}^2}\right)
\vphantom{\sum_{\sigma\,\in\,S_{n+1}}}\right]
\eeqa
with $J_a=x_aP_a$. Notice that each ordering has been assigned its own
$k$-factor.

We further detail the ordered $k$-factor similar to \Eq{KfactorDef} by
dividing the virtual contribution into two parts, which we call
$v_\trm{\tiny D}$ and $f$. The former is proportional to the LO term
and contains the singularities; the latter describes the finite
virtual corrections. We now furthermore specify helicity dependent
$k$-factors and write representative for all orderings
\beqa\label{eq:orderedK}
\lefteqn{
k_\trm{\tiny NLO}\big(J_a^{\lambda_a},J_b^{\lambda_b},
J_1^{\lambda_1},\ldots,J_n^{\lambda_n}\big)\;\,=\;\,
1\;+\;\left(\frac{\alpha_S N_\mrm{C}}{2\pi}\right)\times}
\nnb\\[3mm]&&
\left(v_\trm{\tiny D}\big(J_a,J_b,J_1,\ldots,J_n\big)\;+\;
r\big(J_a,J_b,J_1,\ldots,J_n\big)\;+\;
\hat f\big(J_a^{\lambda_a},J_b^{\lambda_b},
J_1^{\lambda_1},\ldots,J_n^{\lambda_n}\big)\right)\ ,
\eeqa
where
\beq
\hat f\big(J_a^{\lambda_a},J_b^{\lambda_b},
J_1^{\lambda_1},\ldots,J_n^{\lambda_n}\big)\;=\;
\frac{f\big(J_a^{\lambda_a},J_b^{\lambda_b},
  J_1^{\lambda_1},\ldots,J_n^{\lambda_n}\big)}
     {\left|m^{(0)}\big(J_a^{\lambda_a},J_b^{\lambda_b},
       J_1^{\lambda_1},\ldots,J_n^{\lambda_n}\big)\right|^2}\ .
\eeq
Note that the $v_\trm{\tiny D}$ part is independent of the helicity
structure and is known in analytical form~\cite{Giele:1991vf,Giele:1993dj}.
The helicity information is carried by the LO amplitude. We use the
code of Ref.~\cite{Winter:2009kd} to calculate the finite, helicity
dependent contribution $f$. The unresolved soft/collinear
contributions contained in $r$\/ are obtained using a phase-space
slicing method~\cite{Giele:1991vf,Giele:1993dj}. Similarly to the
singular virtual terms, this analytically calculated contribution is
helicity independent and proportional to the Born level expression. We
add it to the $v_\trm{\tiny D}$ term, which now becomes dependent on
the slicing parameter: $v_\trm{\tiny D}\to v_\trm{\tiny D}(s_\mrm{min})$,
while the resolved contribution is contained in $r\to r(s_\mrm{min})$.
The ordered resolved bremsstrahlung factor $r(s_\mrm{min})$ 
is defined such that it is independent of the helicity choice. This is
done in the same manner as for the unordered bremsstrahlung factor
$R=\frac{2\pi\,\widetilde R}{\alpha_S N_\mrm{C}}$ in~\Eq{RDef}. We
define the bremsstrahlung factor as the ratio of the helicity summed
bremsstrahlung contribution divided by the helicity summed LO
contribution, schematically written as
$|m_\trm{brems}^{\lambda_1,\ldots,\lambda_n}|^2\equiv\big(
\sum_\kappa|m_\trm{brems}^{\kappa_1,\ldots,\kappa_{n+1}}|^2/
\sum_\kappa|m_\trm{\tiny LO}^{\kappa_1,\ldots,\kappa_n}|^2\big)
\times |m_\trm{\tiny LO}^{\lambda_1,\ldots,\lambda_n}|^2$. By doing so
we have arranged for having all the helicity dependence carried by the
jets $J_1,\ldots,J_n$, \ie the Born level ordered amplitude. This
approach of defining the helicity dependence gives us the correct
behavior: the soft/collinear limit is found to be proportional to the
LO helicity dependent ordered amplitude. Furthermore, if the helicity
sum over the jets is carried out, one retrieves the helicity summed
bremsstrahlung amplitude.

At this level of the Monte Carlo program development, it is convenient
to have an explicit $\log^2(s_\mrm{min})$ dependence in the unresolved
parton contribution. In the previous section we validated the FBPS
generator based on and using flat phase-space generation. However, the
soft/collinear limit is hardly probed by distributing the momenta
uniformly in phase space. As the $s_\mrm{min}$ dependence of the
$v_\trm{\tiny D}$ part has to cancel against that of the FBPS
generator, we get an excellent probe on the crucial correctness of the
soft/collinear behavior of the FBPS generator.

In the future we can switch to a subtraction method, eliminating
the dependence on the slicing parameter. For the class of infra-red
safe jet algorithms, this is almost trivial. The observable jet final
state is invariant under the bremsstrahlung Monte Carlo integration.
This means we only have to add to the $k$-factor, an integral over the
unresolved phase space. The integrand is simply given by the
difference between the bremsstrahlung matrix element squared and its
antenna approximation. The $s_\mrm{min}$ parameter then becomes
equivalent to the so-called $\alpha$\/ parameter introduced in
Ref.~\cite{Nagy:1998bb}.

\subsection{Numerical studies of NLO high multiplicity jet events}
\label{subsec:jet-numerics}

\begin{figure}[p!]
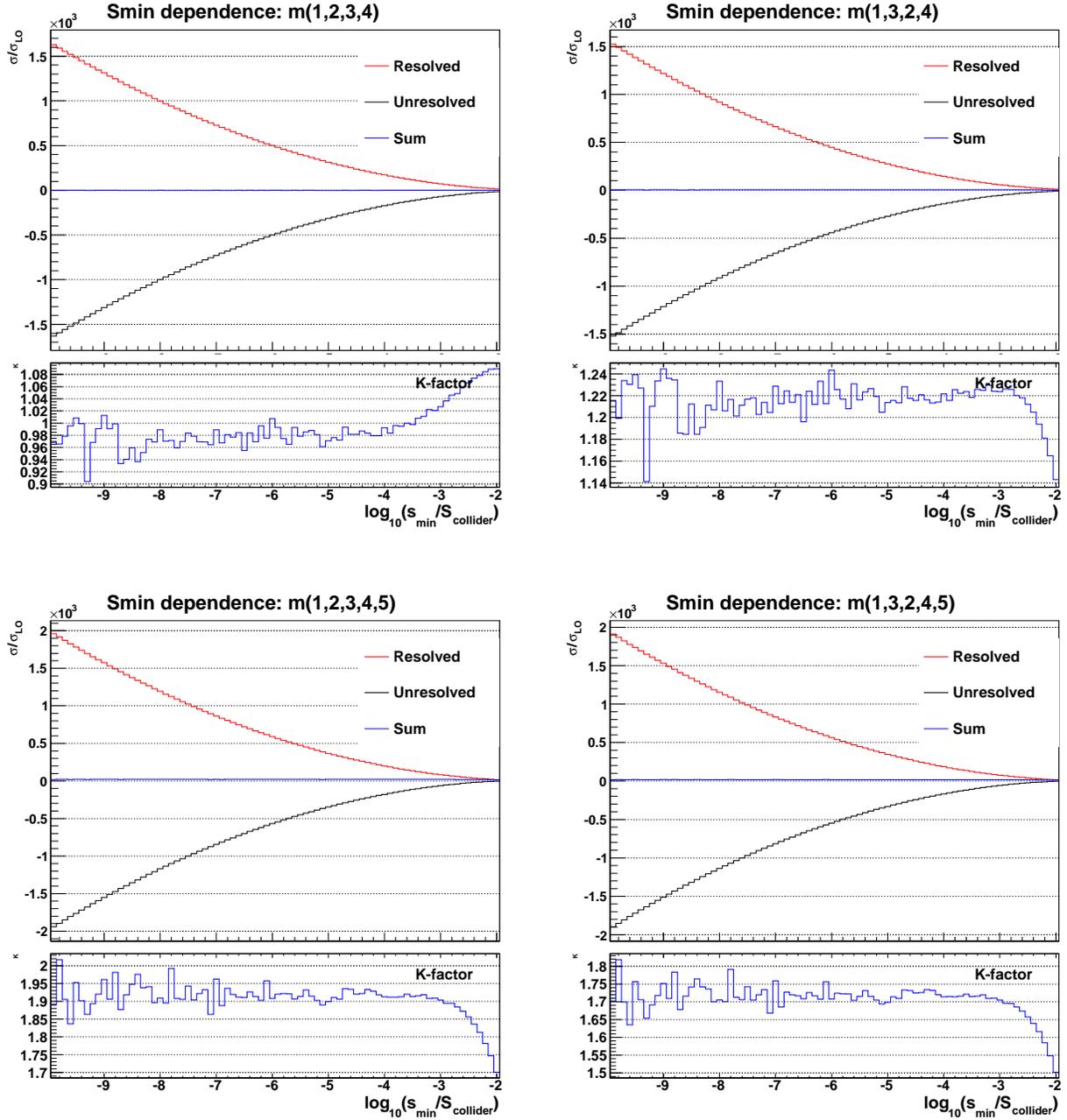

  \centerline{
    \includegraphics[clip,width=0.49\columnwidth,angle=-90]{graphs/%
      Smin2_n10000_b1000_p2.epsi}\hskip10mm
    \includegraphics[clip,width=0.49\columnwidth,angle=-90]{graphs/%
      Smin2_n10000_b1000_p3.epsi}}\vskip10mm
  \centerline{
    \includegraphics[clip,width=0.49\columnwidth,angle=-90]{graphs/%
      Smin3_n10000_b1000_p2.epsi}\hskip10mm
    \includegraphics[clip,width=0.49\columnwidth,angle=-90]{graphs/%
      Smin3_n10000_b1000_p3.epsi}}\vskip5mm
  \caption{\label{smin23}
    The cancellation of the $s_\mrm{min}$ dependence for the 2-jet and
    3-jet configurations given in \App{appendix:explicit}. The upper
    panels show the resolved contribution in red, the unresolved
    contribution in black and the helicity independent part of the
    $k$-factor defined in \Eq{eq:orderedK} in blue. In the lower
    panel, the dependence on $s_\mrm{min}$ of the helicity independent
    part of the $k$-factor is shown in finer detail. The different
    color orderings (indicated in the respective plot titles) use the
    same events in the Monte Carlo evaluation, leading to correlated
    statistical fluctuations.}
\end{figure}

For all numerical studies we use single, exclusive $n$-jet events. As
for the numerical validation of the FBPS generator in
\Sec{subsec:PS-numerics}, all events pass the jet cuts
$p^{(i)}_T>250$~GeV, $|\,\!y_i|<2.0$ and $\Delta R_{ij}>0.5$ at a
collider energy of 7~TeV using the CTEQ6M PDF
set~\cite{Pumplin:2002vw}. The renormalization/factorization scale is
set to one half times the average dijet mass. This scale choice is
closely connected to shower Monte Carlo approaches where the dijet
mass is often used as the starting scale for branchings off the
particular dipole antenna.

\begin{figure}[p!]
  \centerline{
    \includegraphics[clip,width=0.49\columnwidth,angle=-90]{graphs/%
      Smin4_n10000_b1000_p2.epsi}\hskip10mm
    \includegraphics[clip,width=0.49\columnwidth,angle=-90]{graphs/%
      Smin4_n10000_b1000_p4.epsi}}\vskip10mm
  \centerline{
    \includegraphics[clip,width=0.49\columnwidth,angle=-90]{graphs/%
      Smin5_n10000_b1000_p2.epsi}\hskip10mm
    \includegraphics[clip,width=0.49\columnwidth,angle=-90]{graphs/%
      Smin5_n10000_b1000_p6.eps}}\vskip5mm
  \caption{\label{smin45}
    The cancellation of the $s_\mrm{min}$ dependence for the 4-jet and
    5-jet configurations given in \App{appendix:explicit}. The upper
    panels show the resolved contribution in red, the unresolved
    contribution in black and the helicity independent part of the
    $k$-factor defined in \Eq{eq:orderedK} in blue. In the lower
    panel, the dependence on $s_\mrm{min}$ of the helicity independent
    part of the $k$-factor is shown in finer detail. The different
    color orderings (indicated in the respective plot titles) use the
    same events in the Monte Carlo evaluation, leading to correlated
    statistical fluctuations.}
\end{figure}

\begin{figure}[t!]
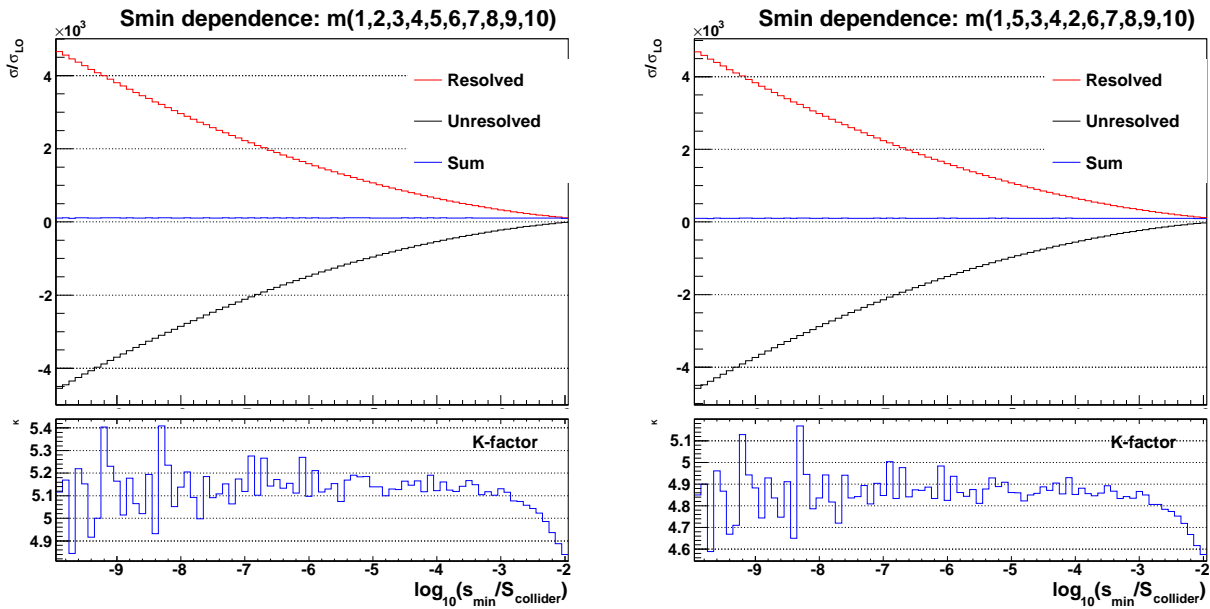

  \vskip5mm
  \centerline{
    \includegraphics[clip,width=0.49\columnwidth,angle=-90]{graphs/%
      Smin8_n10000_b1000_p2.epsi}\hskip10mm
    \includegraphics[clip,width=0.49\columnwidth,angle=-90]{graphs/%
      Smin8_n10000_b1000_p5.epsi}}
  \caption{\label{smin8}
    The cancellation of the $s_\mrm{min}$ dependence for the 8-jet
    configurations given in \App{appendix:explicit}. The upper
    panels show the resolved contribution in red, the unresolved
    contribution in black and the helicity independent part of the
    $k$-factor defined in \Eq{eq:orderedK} in blue. In the lower
    panel, the dependence on $s_\mrm{min}$ of the helicity independent
    part of the $k$-factor is shown in finer detail. The different
    color orderings (indicated in the respective plot titles) use the
    same events in the Monte Carlo evaluation, leading to correlated
    statistical fluctuations.}
\end{figure}

We first have to study the dependence on the slicing parameter
$s_\mrm{min}$ for explicit jet configurations. To this end we
calculate the resolved, $r(s_\mrm{min})$, and the unresolved,
$v_\trm{\tiny D}(s_\mrm{min})$, helicity independent contributions to
the $k$-factor. The results are shown in
\Figs{smin23},~\ref{smin45}~and~\ref{smin8} together with the helicity
independent part of the $k$-factor,
$1+\left(\frac{\alpha_S N_\mrm{C}}{2\pi}\right)
\big[r(s_\mrm{min})+v_\trm{\tiny D}(s_\mrm{min})\big]$, for a single
$2$-jet, $3$-jet, $4$-jet, $5$-jet and $8$-jet event and two different
orderings per kinematic configuration. The notation in the figures is
such that $m^{(0)}(1,2,\ldots,n+2)\equiv m^{(0)}(J_a,J_b,J_1,\ldots,J_n)$
with $1\leftrightarrow a$, $2\leftrightarrow b$, $3\leftrightarrow1$
etc. For each antenna, we only have to perform a three-parameter
integral, giving us good control over the cancellations. The graphs
demonstrate that the cancellation of the $s_\mrm{min}$ parameter
dependence is achieved in a satisfactory manner at values of the order
of $10^{-4}\times S$\/ and smaller. Moreover, we were able to maintain
good numerical stability down to values of $s_\mrm{min}<10^{-9}\times S$
($=S_\mrm{collider}$ in the figures), even though we did not use any
adaptive Monte Carlo integration such as
\textsc{Vegas}~\cite{Lepage:1980dq} to obtain these
results.\footnote{For future applications, the resulting
  three-parameter integration can readily be optimized by important
  sampling and adaptive stratification.}

\begin{table}[t!]
\begin{center}\vskip4mm\small
\begin{tabular}{|c||c||c|c||c|c||c|c||}\hline
  &  & \multicolumn{2}{c||}{} 
     & \multicolumn{2}{c||}{}
     & \multicolumn{2}{c||}{}\\[-3mm]
  &  & \multicolumn{2}{c||}{$--++\cdots+$} 
     & \multicolumn{2}{c||}{$---+\cdots++$}
     & \multicolumn{2}{c||}{$-+-+\cdots-+$}\\[1.5mm]
\cline{3-8}
  &  &&&&&&\\[-3mm]
jets & $r$-factor &
$\left|m^{(0)}\right|^2$ & $k$-factor & $\left|m^{(0)}\right|^2$ & $k$-factor & $\left|m^{(0)}\right|^2$ & $k$-factor \\[1.5mm]
\hline
  &  &&&&&&\\[-1mm]
2  & 172$\pm$ 1   & 1.72216 & 1.15$\pm$ 0.05 & 1.6$\times 10^{-31}$     & $---$          & 0.00552438 & 1.09$\pm$ 0.05 \\[1pt]
3  & 243$\pm$ 2   & 120.638 & 1.13$\pm$ 0.08 & 0.043632                 & 1.18$\pm$ 0.08 & 5.98249    & 1.10$\pm$ 0.08 \\[1pt]
4  & 392$\pm$ 3   & 125.234 & 1.30$\pm$ 0.13 & 0.282847                 & 1.17$\pm$ 0.13 & 0.0498892  & 1.18$\pm$ 0.13 \\[1pt]
5  & 366$\pm$ 4   & 5941.55 & 0.94$\pm$ 0.17 & 849.054                  & 0.87$\pm$ 0.17 & 31.5083    & 0.80$\pm$ 0.17 \\[1pt]
6  & 529$\pm$ 5   & 1202.54 & 1.15$\pm$ 0.24 & 69.0066                  & 1.06$\pm$ 0.24 & 0.469815   & 0.82$\pm$ 0.24 \\[1pt]
8  & 650$\pm$ 7   & 26732.0 & 1.41$\pm$ 0.34 & 1364.49                  & 1.32$\pm$ 0.34 & 1.41604    & 1.15$\pm$ 0.34 \\[1pt]
10 & 844$\pm$ 11  & 6575.23 & 1.49$\pm$ 0.49 & 579.066         & 1.26$\pm$ 0.49 & 6.09232$\times 10^{-6}$ & 0.97$\pm$ 0.49 \\[1pt]
15 & 1264$\pm$20  & 4690.02 & 1.39$\pm$ 0.95 & 671.554         & 1.28$\pm$ 0.95 & 4.37178$\times 10^{-7}$ & 1.24$\pm$ 0.95 \\[3mm]\hline
\end{tabular}
\end{center}
\caption{\label{X-sections}
  The LO ordered amplitude squared $|m^{(0)}(J_a,J_b,J_1,\ldots,J_n)|^2$
  and its corresponding $r(s_\mrm{min})$ and ordered $k$-factor as
  defined in \Eq{eq:orderedK} for an exclusive $n$-jet event. The
  explicit jet momenta for the different jet multiplicities are given
  in \App{appendix:explicit}. The slicing scale $s_\mrm{min}$ is set
  to $10^{-4}\times S$\/ and the Monte Carlo integration over the
  bremsstrahlung phase space has been done with 100,000 generated
  events.}
\end{table}

We can now proceed to make some predictions for the $n$-jet
configurations listed in \App{appendix:explicit}. In \Tab{X-sections}
we show the results for several helicity configurations of the
calculations of the $k$-factors defined in \Eq{eq:orderedK}
multiplying the ordered amplitudes squared. As can be seen from the
table, the scale choice described above leads to relatively small
$k$-factors, \ie the normalization of the LO prediction is fairly
close to the NLO rate. In other words, the correction of the LO weight
due to radiative corrections is of the order of one.
In addition to the $k$-factors, our table also displays
the numerical results $r(s_\mrm{min})$ of the bremsstrahlung
phase-space integration using the FBPS generator. As expected, for a
fixed number of bremsstrahlung events, the uncertainty on the
integration results scales with the square root of the number of
dipoles, \ie it grows linearly in the number of final-state
jets (except for the two highest jet multiplicities, where with the
chosen number of events we did not achieve a sufficient accuracy for
determining a reliable uncertainty estimate). The number of
final--final and initial--final dipoles is $n(n-1)$ and $2n$,
respectively. To obtain an integration uncertainty independent of the
$n$-jet multiplicity, we would have to scale the number of
bremsstrahlung events evaluated in the Monte Carlo integration as
$n(n-1)+2n=n(n+1)$.

From \Tab{X-sections} one also reads off that the relative errors on
the $k$-factors become sizeable for a large number of jets. This is
when the values of the $r$-factors turn big and one finds large
cancellations between the single terms in \Eq{eq:orderedK}. Owing to
the use of a slicing method and its explicit dependence on
$\log(s_\mrm{min})$, we are not able to avoid this behavior easily
without switching to an approach based on subtractions. Because of
these large cancellations, the absolute values of $\Delta r$\/
multiplied by $\left(\frac{\alpha_S N_\mrm{C}}{2\pi}\right)\sim1/20$
are hence pivotal in determining the uncertainty on the $k$-factors.


\section{Conclusions}\label{sec:Conclusions}

In this paper we derived a new type of NLO phase-space generator.
This forward-branching phase-space generator has the property of
inverting the clustering occurring in the jet algorithm. Because
of this, the bremsstrahlung phase space of a fully exclusive jet
final state is generated. The bremsstrahlung events constructed in
this way do not change the value of a jet observable; they are all
added to the single virtual contribution. This gives a perfect
cancellation of the divergent pieces -- the soft/collinear real and
virtual contributions -- as dictated by the KLN theorem for any jet
observable.

However, the current jet algorithms used by the experiments
employ a $2\to1$ clustering scheme. Furthermore,
a beam jet is not defined leading to transverse momentum imbalance
in jet events. As a result the LO and NLO jet phase
spaces are different and only match at the soft/collinear boundary.
This makes the jet observables infra-red finite, but not necessarily 
infra-red safe. 
The KLN theorem becomes applicable only after some phase-space
averaging. As a consequence, the fully exclusive multi-jet event is not
defined at NLO.

Any infra-red finite jet algorithm using a $3\rightarrow 2$ clustering
scheme, which includes beam jets, suffices to define the fully
exclusive multi-jet differential cross section. For the purposes of
this paper, we worked out a theoretically simple jet algorithm such
that fully exclusive jet final states can be defined at NLO.
Nevertheless, the difference between an observable determined by this
more theoretical jet algorithm and one currently used by the
experiments amounts to a finite correction. This correction is readily
calculated for a particular observable by an extra bremsstrahlung
phase-space integration, extending the applicability of the
forward-branching phase-space generator beyond jet algorithms using
$3\rightarrow 2$ clustering schemes.

Since the forward-branching bremsstrahlung phase-space generator does
not alter the jet configuration, a $K$-factor can be defined for a
given multi-jet phase-space point. The $K$-factor can be determined
from the leading-order probability associated with the particular jet
phase-space point. In this sense, the probabilistic interpretation of
the NLO prediction is restored by quantifying it as a positive-weight
adjustment. We validated our method in two steps: first, we verified
the construction of the bremsstrahlung phase space; second, we
calculated the radiative corrections for individual events with up to
$15$ jets. As our test scenario, we chose $2\to n$\/ gluon production
in the leading-color approximation.


\section*{Acknowledgments}

J.~W.\ thanks the Fermilab Theory Group for the great hospitality
during his visit early in 2010. Financial support is acknowledged.

Fermilab is operated by Fermi Research Alliance, LLC, under contract
DE-AC02-07CH11359 with the United States Department of Energy.


\appendix

\section{The Enumeration of Branchings}\label{appendix:counting}

We can choose the branching pair of jets in \Eq{RDef} by Monte Carlo
means. However, it is useful to control the ratio of initial--final
and final--final branchings for optimization purposes. This ratio is
given by the number of ways the $n+3$ partons can be clustered to
$n$\/ jets plus $2$ incoming partons. If no flavor constraints are
taken into account, the derivation of the ratio for the jet algorithm
used in this paper goes as follows:
\begin{itemize}
\item[--] Given an initial-state parton, there are $n\,(n+1)$ possible
  clusterings since it is the final-state parton pair that effectively
  gets combined. This number is based on treating all final-state
  partons as distinguishable particles. The recoil is taken by one
  of the two partons in the initial state, giving the number of
  initial--final state clusterings
  $N_\trm{\tiny IF}=2\,n\,(n+1)$.
\item[--] The number of final--final state clusterings is given by
  $n\,(n+1)$. For each of these clusterings, we have $n-1$ remaining
  possible recoil partons. For the number of final-state clusterings,
  we hence obtain $N_\trm{\tiny FF}=(n-1)\,n\,(n+1)$.
\item[--] Accordingly, the total number of possible clusterings is
  $N_\mrm{tot}=N_\trm{\tiny IF}+N_\trm{\tiny FF}=n\,(n+1)^2$.
\item[--] Following these observations, the fraction of final--final
  state branchings to be generated is then given by
  $N_\trm{\tiny FF}/N_\mrm{tot}=(n-1)/(n+1)$. Similarly, the fraction
  of initial--final state branchings thus is
  $N_\trm{\tiny IF}/N_\mrm{tot}=2/(n+1)$.
  As we see for a very large number of jets, the fraction of pure
  final-state branchings tends towards one as expected.
\end{itemize}
Note that in \Eq{RDef} we see that each dipole term is ``averaged''
by exactly the number of final--final state dipoles, $N_\trm{\tiny FF}$,
or the number of initial--final state dipoles per beam, 
$\frac{1}{2}N_\trm{\tiny IF}$.

\section{The Explicit Jet Events}\label{appendix:explicit}

The jet configurations used to calculate the results in
\Tab{X-sections} are given in this appendix, together with some
kinematic properties of the particular $n$-jet event. These properties
are calculated from the final-state jets.

\small\begin{itemize}
\item For $n=2$, we have:
  \bea
  \alpha_S\left(\frac{1}{2}\langle m_{jj}\rangle\right)&=&0.0854525\ ,\quad
  \langle m_{jj}\rangle\;=\;2320.1\ \mrm{GeV}\ ,\nnb\\
  \min\left(m_{jj}\right)&=&2320.1\ \mrm{GeV}\ ,\quad
  \max\left(m_{jj}\right)\;=\;2320.1\ \mrm{GeV}
  \eea
  and the momenta (in GeV) are given by
  \bea
  x_a\,P_a&=&(651.429,651.429,0,0)\ ,\nnb\\
  x_b\,P_b&=&(2065.78,-2065.78,0,0)\ ,\nnb\\
  J_1&=&(988.026,4.76957,150.427,976.496)\ ,\nnb\\
  J_2&=&(1729.19,-1419.12,-150.427,-976.496)\ .
  \eea
\item For $n=3$, we have:
  \bea
  \alpha_S\left(\frac{1}{2}\langle m_{jj}\rangle\right)&=&0.08936\ ,\quad
  \langle m_{jj}\rangle\;=\;1546.75\ \mrm{GeV}\ ,\nnb\\
  \min\left(m_{jj}\right)&=&837.178\ \mrm{GeV}\ ,\quad
  \max\left(m_{jj}\right)\;=\;2342.36\ \mrm{GeV}
  \eea
  and the momenta (in GeV) are given by
  \bea
  x_a\,P_a&=&(1002.78,1002.78,0,0)\ ,\nnb\\
  x_b\,P_b&=&(1789.36,-1789.36,0,0)\ ,\nnb\\
  J_1&=&(1203.62,-339.322,1151.26,-90.3834)\ ,\nnb\\
  J_2&=&(1243.44,-297.018,-1187.58,-218.141)\ ,\nnb\\
  J_3&=&(345.076,-150.236,36.3206,308.525)\ .
  \eea
\item For $n=4$, we have:
  \bea
  \alpha_S\left(\frac{1}{2}\langle m_{jj}\rangle\right)&=&0.0873363\ ,\quad
  \langle m_{jj}\rangle\;=\;1899.39\ \mrm{GeV}\ ,\nnb\\
  \min\left(m_{jj}\right)&=&906.006\ \mrm{GeV}\ ,\quad
  \max\left(m_{jj}\right)\;=\;2636.08\ \mrm{GeV}
  \eea
  and the momenta (in GeV) are given by
  \bea
  x_a\,P_a&=&(2293.54,2293.54,0,0)\ ,\nnb\\
  x_b\,P_b&=&(2359.46,-2359.46,0,0)\ ,\nnb\\
  J_1&=&(1725.71,877.743,982.549,1114.55)\ ,\nnb\\
  J_2&=&(954.182,356.832,-669.803,-578.359)\ ,\nnb\\
  J_3&=&(1037.51,-521.634,-809.127,-386.844)\ ,\nnb\\
  J_4&=&(935.596,-778.871,496.381,-149.35)\ .
  \eea
\item For $n=5$, we have:
  \bea
  \alpha_S\left(\frac{1}{2}\langle m_{jj}\rangle\right)&=&0.0921983\ ,\quad
  \langle m_{jj}\rangle\;=\;1074.2\ \mrm{GeV}\ ,\nnb\\
  \min\left(m_{jj}\right)&=&365.996\ \mrm{GeV}\ ,\quad
  \max\left(m_{jj}\right)\;=\;1794.95\ \mrm{GeV}
  \eea
  and the momenta (in GeV) are given by
  \bea
  x_a\,P_a&=&(883.985,883.985,0,0)\ ,\nnb\\
  x_b\,P_b&=&(3263.37,-3263.37,0,0)\ ,\nnb\\
  J_1&=&(684.733,-446.345,-519.146,11.084)\ ,\nnb\\
  J_2&=&(780.483,-90.2618,684.399,-364.149)\ ,\nnb\\
  J_3&=&(1081.8,-949.154,502.292,130.739)\ ,\nnb\\
  J_4&=&(458.187,-330.582,216.105,-232.271)\ ,\nnb\\
  J_5&=&(1142.15,-563.042,-883.651,454.598)\ .
  \eea
\item For $n=6$, we have:
  \bea
  \alpha_S\left(\frac{1}{2}\langle m_{jj}\rangle\right)&=&0.0898744\ ,\quad
  \langle m_{jj}\rangle\;=\;1470.28\ \mrm{GeV}\ ,\nnb\\
  \min\left(m_{jj}\right)&=&372.579\ \mrm{GeV}\ ,\quad
  \max\left(m_{jj}\right)\;=\;2307.33\ \mrm{GeV}
  \eea
  and the momenta (in GeV) are given by
  \bea
  x_a\,P_a&=&(2711.42,2711.42,0,0)\ ,\nnb\\
  x_b\,P_b&=&(2989.73,-2989.73,0,0)\ ,\nnb\\
  J_1&=&(1305.48,-936.767,-465.008,781.36)\ ,\nnb\\
  J_2&=&(416.233,43.8532,163.26,-380.359)\ ,\nnb\\
  J_3&=&(1255.05,418.867,211.911,1163.96)\ ,\nnb\\
  J_4&=&(809.408,-755.773,-267.267,-111.885)\ ,\nnb\\
  J_5&=&(1029.43,606.551,515.234,-652.956)\ ,\nnb\\
  J_6&=&(885.548,344.959,-158.13,-800.121)\ .
  \eea
\item For $n=8$, we have:
  \bea
  \alpha_S\left(\frac{1}{2}\langle m_{jj}\rangle\right)&=&0.0956911\ ,\quad
  \langle m_{jj}\rangle\;=\;861.375\ \mrm{GeV}\ ,\nnb\\ 
  \min\left(m_{jj}\right)&=&239.955\ \mrm{GeV}\ ,\quad
  \max\left(m_{jj}\right)\;=\;1483.52\ \mrm{GeV}
  \eea
  and the momenta (in GeV) are given by
  \bea
  x_a\,P_a&=&(1566.71,1566.71,0,0)\ ,\nnb\\
  x_b\,P_b&=&(3315.09,-3315.09,0,0)\ ,\nnb\\
  J_1&=&(771.638,72.2583,756.755,132.381)\ ,\nnb\\
  J_2&=&(799.106,-447.319,-526.283,-401.875)\ ,\nnb\\
  J_3&=&(362.664,-225.218,127.162,-254.228)\ ,\nnb\\
  J_4&=&(455.86,294.142,-344.111,-53.627)\ ,\nnb\\
  J_5&=&(336.571,-176.658,48.6816,282.315)\ ,\nnb\\
  J_6&=&(897.978,-705.684,548.115,89.1335)\ ,\nnb\\
  J_7&=&(897.601,-802.832,-399.01,44.0329)\ ,\nnb\\
  J_8&=&(360.374,242.933,-211.31,161.867)\ .
  \eea
\item For $n=10$, we have:
  \bea
  \alpha_S\left(\frac{1}{2}\langle m_{jj}\rangle\right)&=&0.0950109\ ,\quad
  \langle m_{jj}\rangle\;=\;913.794\ \mrm{GeV}\ ,\nnb\\
  \min\left(m_{jj}\right)&=&237.586\ \mrm{GeV}\ ,\quad
  \max\left(m_{jj}\right)\;=\;1830.56\ \mrm{GeV}
  \eea
  and the momenta (in GeV) are given by
  \bea
  x_a\,P_a&=&(2827.46,2827.46,0,0)\ ,\nnb\\
  x_b\,P_b&=&(3322.41,-3322.41,0,0)\ ,\nnb\\
  J_1&=&(547.589,302.187,339.049,305.912)\ ,\nnb\\
  J_2&=&(956.324,-806.585,389.162,335.454)\ ,\nnb\\
  J_3&=&(350.22,-95.7603,228.577,247.461)\ ,\nnb\\
  J_4&=&(259.829,-26.1444,-35.2505,-256.096)\ ,\nnb\\
  J_5&=&(314.102,-35.8975,4.96171,312.004)\ ,\nnb\\
  J_6&=&(986.504,-606.912,589.914,-506.804)\ ,\nnb\\
  J_7&=&(889.429,-326.063,-798.032,-218.89)\ ,\nnb\\
  J_8&=&(891.285,649.951,-560.226,-241.04)\ ,\nnb\\
  J_9&=&(518.799,168.343,-476.506,117.284)\ ,\nnb\\
  J_{10}&=&(435.784,281.925,318.35,-95.2848)\ .
  \eea
\item For $n=15$, we have:
  \bea
  \alpha_S\left(\frac{1}{2}\langle m_{jj}\rangle\right)&=&0.0993949\ ,\quad
  \langle m_{jj}\rangle\;=\;633.545\ \mrm{GeV}\ ,\nnb\\
  \min\left(m_{jj}\right)&=&147.1\ \mrm{GeV}\ ,\quad
  \max\left(m_{jj}\right)\;=\;1497.81\ \mrm{GeV}
  \eea
  and the momenta (in GeV) are given by
  \bea
  x_a\,P_a&=&(3403.1,3403.1,0,0)\ ,\nnb\\
  x_b\,P_b&=&(3096.06,-3096.06,0,0)\ ,\nnb\\
  J_1&=&(349.237,141.394,-212.337,-238.51)\ ,\nnb\\
  J_2&=&(272.807,-73.295,214.113,-152.339)\ ,\nnb\\
  J_3&=&(465.556,289.402,279.374,234.392)\ ,\nnb\\
  J_4&=&(269.21,54.8123,-263.571,-0.174956)\ ,\nnb\\
  J_5&=&(451.568,254.001,87.8387,-362.88)\ ,\nnb\\
  J_6&=&(437.866,292.297,303.293,-119.594)\ ,\nnb\\
  J_7&=&(368.167,-241.648,223.073,165.504)\ ,\nnb\\
  J_8&=&(355.366,-41.1757,-217.718,-277.828)\ ,\nnb\\
  J_9&=&(328.823,-87.4318,-296.434,-112.281)\ ,\nnb\\
  J_{10}&=&(458.585,-188.764,-88.852,408.379)\ ,\nnb\\
  J_{11}&=&(332.789,159.417,-133.945,259.603)\ ,\nnb\\
  J_{12}&=&(572.199,-470.06,-321.957,-52.8996)\ ,\nnb\\
  J_{13}&=&(746.136,-332.113,245.182,-621.534)\ ,\nnb\\
  J_{14}&=&(830.982,518.077,212.244,614.069)\ ,\nnb\\
  J_{15}&=&(259.874,32.133,-30.3028,256.093)\ .
  \eea
\end{itemize}
\bigskip\normalsize


\bibliographystyle{JHEP}
{\raggedright\bibliography{fbps}}
\vspace*{0mm}\noindent\hrulefill


\end{document}